\documentclass[12pt, uplatex]{article}
\usepackage{amsmath,amssymb}
\usepackage{graphicx}
\usepackage{bm}
\usepackage{color}
\usepackage{amscd}
\usepackage{slashed}
\usepackage[left=2cm,right=2cm,top=2.5cm,bottom=3cm]{geometry}
\usepackage{titling}

\newcommand{\nn}{\nonumber }
\newcommand{\vk}{\varkappa}
\newcommand{\rmax}{r_\text{max}}
\newcommand{\rmaxz}{r_\text{max,0}}

\allowdisplaybreaks

\begin{document}

\title{Spin chains and classical strings in two parameters $q$-deformed AdS$_3\times$S$^3$}

\author{
	Wen-Yu Wen$^{*\dagger}$
        and
        Shoichi Kawamoto$^*$ \bigskip\\
\footnotesize\tt %kawamoto@yukawa.kyoto-u.ac.jp,
wenw@cycu.edu.tw,
kawamoto@cycu.edu.tw
	\\
	$^*$\footnotesize\it 
Center for High Energy Physics and
Department of Physics,\\
\footnotesize\it 
 Chung-Yuan Christian University,
Chung-Li 320, Taiwan, R.O.C.\\
	$^\dagger$\footnotesize\it 
Leung Center for Cosmology and Particle Astrophysics,
National Taiwan University, Taipei 106, Taiwan
	}

% \author{Shoichi Kawamoto}\thanks{%
% E-mail: kawamoto@cycu.edu.tw, kawamoto@yukawa.kyoto-u.ac.jp
% }
% \affiliation{Department of Physics and Center for High Energy Physics, Chung Yuan Christian University, Chung Li District, Taiwan}

% \author{Shogo Kuwakino}\thanks{%
% }
% \affiliation{Department of Physics and Center for High Energy Physics, Chung Yuan Christian University, Chung Li District, Taiwan}
% \author{Wen-Yu Wen}\thanks{%
% E-mail: steve.wen@gmail.com}
% \affiliation{Department of Physics and Center for High Energy Physics, Chung Yuan Christian University, Chung Li District, Taiwan}  
% \affiliation{Leung Center for Cosmology and Particle Astrophysics\\
% National Taiwan University, Taipei 106, Taiwan}

%\preprint{CYCU-HEP-19-08}

\bigskip

\begin{titlingpage}

    \maketitle

\begin{abstract}
\noindent %\normalsize

In this paper, we study the spin chain and string excitation in the two-parameters $q$-deformed AdS$_3\times$S$^3$ proposed by Hoare \cite{Hoare:2014oua}.  We obtain the deformed spin chain model at the fast spin limit for choices of deformed parameters.  General ansatz for giant magnons are studied in great detail and complicated dispersion relation is treated perturbatively. We also study several types of hanging string solutions and their charges and spins are analyzed numerically.  At last, we explore its pp-wave limit and find its solution only depends on the difference of deformed parameters.
\end{abstract}
\end{titlingpage}

%\pacs{04.70.Dy    04.70.-s    04.62.+v}
%\maketitle

%\date{\today}

%\keywords{}

\clearpage

\section{Introduction and summary}

Since AdS/CFT correspondence was first advocated two decades ago \cite{Maldacena:1997re},
the dictionary of correspondence has been getting more elaborated and precise.
One of the sharpest correspondence has been observed in the correspondence between
a single trace operator of large
dimension in  ${\cal N}=4$ super Yang-Mills theory in large-$N$ limit
and a sting/brane configuration in AdS${}_5\times S^5$ spacetime \cite{integ_big_review}.
Integrability has played a prominent role in this correspondence; the problem of finding anomalous dimension of the single trace operator leads to a problem of an integrable spin chain, and the sigma model on the string world sheet also exhibits integrability.
The sigma model may be deformed with integrable structure intact
and this deformation would correspond to a choice of new background that preserves 
integrability.
This integrable deformation has been attracting attention and stimulates many works
\cite{integ_big_review,Swanson:2007dh}.
Among many works that focus on one parameter deformation of the sigma model,
Hoare proposed a two-parameter deformation of sigma models on the world-sheet of superstrings
on AdS$_3 \times S^3 \times M^4$ with $M^4$ being $T^4$ or $S^1 \times S^3$ \cite{Hoare:2014oua},
and a deformed metric on AdS$_3 \times S^3$ was also presented based on the 
sigma-model deformation.
In this paper, we consider some string solutions on this deformed background
and analyze the dispersion relation.
With this new deformed metric, we are interested in investigating several types of
string solution in this background.

The deformed metric involves two deformation parameter, $\vk_+$ and $\vk_-$,
and we can assume $\vk_+ \geq \vk_-$ without loss of generality due to the symmetry $\vk_+ \leftrightarrow
\vk_-$.
By taking $\vk_- \rightarrow 0$ with $\vk_+$ fixed, it becomes a one parameter deformation
case \cite{Arutyunov:2013ega}, while $\vk_- \rightarrow \vk_+$ limit provides a known  squashed $S^3$ case \cite{Cherednik:1981df}.
We thus explore the effect of the two parameter deformation near these special point.
The dispersion relations turn out to be complicated but some explicit forms are to be given
as a perturbative series.

The organization of the paper is as follows:
In Section \ref{sec:spin-chain-from}, the spin chain Hamiltonian
obtained from the sigma model on this back ground is discussed.
In Section \ref{sec:giant-magnon-spiky},
long string solutions in this background and its dispersion relation
is analyzed.
The PP-wave limit is briefly discussed in Section \ref{sec:pp-wave-limit}.
Section \ref{sec:conlusion} serves summary and discussion.
The appendix \ref{sec:two-param-deform} summarizes some facts on the two-parameter deformed geometry.

\section{Spin chain from $q^2$-deformed AdS$_3$ (S$^3$)}
\label{sec:spin-chain-from}

The Heisenberg spin chain could be derived from the sigma model in the fast spinning limit, and its Hamiltonian agrees with the one-loop calculation of anomalous dimensions in ${\cal N}=4$ super Yang-Mills theory \cite{Kruczenski:2003gt}.  Although this quantity may no longer be protected in the deformed theory with less symmetry, one can still study the effect of deformation to the spin chain Hamiltonian and dynamics from gravity side.  Let us consider the deformed metric in $R\times S^3 \in AdS_3\times S^3$\cite{Hoare:2014oua}.
% \begin{align}
% ds^2 =&\frac{1}{1+\kappa_-^2(1+\tilde{\rho}^2)-\kappa_+^2\tilde{\rho}^2}\big[ -(1+\tilde{\rho}^2)(1+\kappa_-^2(1+\tilde{\rho}^2))dt^2 \big]
% \nn\\&
% + \frac{1}{1+\kappa_-^2(1-r^2)+\kappa_+^2r^2}\big[ \frac{dr^2}{1-r^2}
% +(1-r^2)(1-\kappa_-^2(1-r^2)d\varphi^2+r^2(1+\kappa_+^2r^2)d\phi^2
% \nn\\& \hskip8em
% +2\kappa_+\kappa_-r^2(1-r^2)d\varphi d\phi \big]
% \end{align}
% with a change of variable $\tilde{\rho}=\sinh \rho$ and $r = \sin\theta$, the above metric becomes
% \begin{align}
% \label{theta_metric}
% ds^2 =&\frac{1}{1+\kappa_-^2\cosh^2\rho-\kappa_+^2\sinh^2\rho}\big[ -\cosh^2\rho(1+\kappa_-^2\cosh^2\rho)dt^2 \big]
% \nn\\&
% +\frac{1}{1+\vk_-^2\cos^2\theta+\vk_+^2\sin^2\theta}
% \bigg[
% d\theta^2
% +\cos^2\theta \big( 1+\vk_-^2\cos^2\theta \big) d\varphi^2
% +\sin^2\theta (1+\vk_+^2 \sin^2\theta)d\phi^2
% \nn\\& \hskip4em
% +2\vk_-\vk_+\sin^2\theta \cos^2\theta d\varphi d\phi
% \bigg] \,.
% \end{align}
In the deformed coordinates \eqref{eq:A3} and \eqref{eq:S3},
after a twist of angular coordinates $\varphi=\phi_1+\phi_2$ and $\phi=\phi_1-\phi_2$ for simplicity, we consider a spinning string at the center of $AdS$ by rotating coordinate $\phi_1\to t+\tilde{\phi_1}$ and setting $\rho=0$.  The metric then reads
\begin{align}
  ds^2 =&
\frac{1}{1+\vk_-^2\cos^2\theta+\vk_+^2\sin^2\theta}
\bigg[
d\theta^2
-(\vk_+-\vk_-)^2 \cos^2\theta \sin^2\theta \, dt^2
\nn\\& \hskip3em
+\big\{
1+(\vk_- \cos^2\theta + \vk_+ \sin^2\theta)^2
\big\} \big( 2dtd\tilde{\phi}_1 + d\tilde{\phi}_1^2  \big)
\nn\\& \hskip3em
+\big\{
1+(\vk_- \cos^2\theta - \vk_+ \sin^2\theta)^2
\big\} d\phi_2^2 
\nn\\& \hskip3em
+\big(\cos 2\theta + \vk_-^2 \cos^4\theta - \vk_+^2 \sin^4\theta \big) 
\big(2 dt d\phi_2 + 2 d\tilde{\phi}_1 d\phi_2 \big)
\bigg] \,.
\end{align}

% %the metric reads
% %\begin{eqnarray}
% %ds^2 &=& -dt^2 +\frac{1}{1-\kappa_-^2\cos^2{\theta}+\kappa_+^2\sin^2{\theta}}\big[ d\theta^2 + d\phi_1^2 + d\phi_2^2 + (\cos^2\theta -\kappa_-^2\cos^4\theta-\kappa_+^2\sin^4\theta)(2d\phi_1d\phi_2)\nonumber\\
% %&-&(\kappa_-^2\cos^4\theta-2\kappa_+\kappa_-\sin^2\theta\cos^2\theta-\kappa_+^2\sin^4\theta)d\phi_1^2-(\kappa_-^2\cos^4\theta+2\kappa_+\kappa_-\sin^2\theta\cos^2\theta-\kappa_+^2\sin^4\theta)d\phi_2^2 \big]
% %\big]
% %\end{eqnarray}

% \begin{eqnarray}
% ds^2 &=& \frac{1}{1-\kappa_-^2\cos^2\theta+\kappa_+^2\sin^2\theta}\big[  d\theta^2 + (1-\kappa_-^2\cos^4\theta+2\kappa_+\kappa_-\sin^2\theta\cos^2\theta+\kappa_+^2\sin^4\theta)(2dtd\tilde{\phi_1}) \nonumber\\
% &+& (\kappa_-^2+2\kappa_+\kappa_--\kappa_+^2)\sin^2\theta\cos^2\theta dt^2 + (\cos{2\theta}-\kappa_-^2\cos^4\theta-\kappa_+^2\sin^4\theta)(2dtd\phi_2)\nonumber\\ 
% &+&(1-\kappa_-^2\cos^4\theta+2\kappa_+\kappa_-\sin^2\theta\cos^2\theta+\kappa_+^2\sin^4\theta)d\tilde{\phi_1}^2 \nonumber\\
% &+& (1-\kappa_-^2\cos^4\theta-2\kappa_+\kappa_-\sin^2\theta\cos^2\theta+\kappa_+^2\sin^4\theta)d\phi_2^2 \nonumber\\
% &+& (\cos{2\theta}-\kappa_-^2\cos^4\theta-\kappa_+^2\sin^4\theta)2d\tilde{\phi_1}d\phi_2 \big]
% \end{eqnarray}
The case of one-parameter deformation, i.e. $\vk_-=0$, has been studied in \cite{Kameyama:2014bua}.  Here we will focus on those novel cases where $\vk_- \neq 0$.  With the choice of gauge $t=\kappa_0 \tau$ and taking fast-moving limit: $\vk_\pm \to 0$, $\dot{X^{\mu}}\to 0$, $\kappa_0 \to \infty$, but $\kappa_0\vk_\pm$ and $\kappa_0 \dot{X^\mu}$ being kept fixed, one obtains the pull-back string action:
\begin{equation} 
S = \frac{T}{2} \int d\tau d\sigma \, \bigg[
C%\vk_-^2\kappa_0^2
 \sin^2{2\theta} + 2\kappa_0 (\dot{\tilde{\phi_1}}+\cos{2\theta}\dot{\phi_2})-{\theta^\prime}^2 -{\tilde\phi_1^{\prime 2}}-\phi_2^{\prime 2}
-2\cos{2\theta}{\tilde\phi_1}^\prime\phi_2^\prime \bigg] \,,
\end{equation}
where $C= -{\kappa_0^2(\vk_+-\vk_-)^2}/{4}$.
Using one of the Virasoro constraints,
\begin{equation}
2\kappa_0 ({\tilde\phi_1}'+\cos{2\theta}\phi_2^\prime)=0,
\end{equation}
one obtains a $q-$deformed spin chain with Hamiltonian density
\begin{equation}
H = \theta^{\prime 2} + \sin^2{2\theta} \phi_2^{\prime 2} - C \sin^2{2\theta}
\end{equation}
This is more or less a Heisenberg XXZ spin chain.  To see that, one first rescales $\theta \to \theta/2$, $\phi_2 \to \phi_2/2$ and defines a new Hamiltonian ${\cal H}\equiv 4(H+C)$.  Then one obtains a spin chain interacting with magnetic field along z-axis, that is
\begin{equation}
{\cal H} = \vec{n}^{\prime}\cdot \vec{n}^{\prime} + (\vec{n}\cdot\vec{B})^2,
\end{equation}
where the spin vector $\vec{n} = (\sin\theta \cos\phi_2, \sin\theta\sin\phi_2, \cos\theta)$ and an external magnetic field $\vec{B}=(0,0,2\sqrt{|C|})$.  Similar result was obtained before in the case of deformed three-sphere\cite{Wen:2006fw}.

% \section{Two-spin strings}

% \subsection{GKP}

\section{Open string solutions}
\label{sec:giant-magnon-spiky}

\subsection{One-spin giant magnon solution}
\label{sec:one-spin-giant}

We start with a simple case of the basic giant magnon solution
that has one spin in $S^3$, 
by following \cite{Khouchen:2014kaa}.
Because of $\mathbf{Z}_2$ symmetry of $\vk_+ \leftrightarrow \vk_-$, we assume
$\vk_+ \geq \vk_-$ without loss of generality;
the $\vk_- \rightarrow 0$ limit corresponds to
the one-parameter deformation case.
We take the ansatz
\begin{align}
  t =& \kappa \tau \,,
\qquad
\psi=\Psi \tau \,,
\qquad
\rho=\rho(\tau) \,,\\
\varphi=& \omega ( \tau + h(y) ) \,,
\qquad
\theta = \theta(y) \,,
\qquad
\phi=0 \,,
\end{align}
where $y=\sigma-v \tau$ and $0< v<1$.

The equation of motion for $t$ admits a constant $\rho$ solution.
We choose $\rho=0$ which is consistent with
the equation of motion.
The action and the Virasoro constraints are
\begin{align}
  I=& -\frac{\hat{T}}{2} \int d\tau dy \,
\bigg[\kappa^2
+(1-v^2) g_{\theta\theta} \theta^{\prime 2}
+ \omega^2 g_{\varphi\varphi} \big(
(1-v^2) h^{\prime 2} +2v h' -1 \big)
\bigg]
\,,\\
0=&
-\kappa^2 
+(1+v^2) g_{\theta\theta} \theta^{\prime 2}
+ \omega^2 g_{\varphi\varphi} \big(
(1+v^2) h^{\prime 2} -2v h' +1 \big)
\,,\\
0=&
-v g_{\theta\theta} \theta^{\prime 2}
+ \omega^2 g_{\varphi\varphi} h' ( 1-v h')
\,,
\end{align}
where the prime denotes $y$ derivative.
The $h$ equation, once integrated, leads to
\begin{align}
  \omega^2 g_{\varphi\varphi} \big(
(1-v^2) h^{\prime} +v \big)  = C \,,
\end{align}
where $C$ is the constant of integration.
By eliminating $\theta^{\prime 2}$ from two Virasoro constraints,
one finds $C= v \kappa^2$.
Thus,
\begin{align}
  h' =& \frac{v}{1-v^2} \bigg[
\frac{\kappa^2}{\omega^2 g_{\varphi\varphi}}
-1
\bigg] \,.
\end{align}
The $\theta$ equation is
\begin{align}
  \theta^{\prime 2} =&
\frac{\omega^2}{g_{\theta\theta} g_{\varphi\varphi}(1-v^2)^2}
\bigg( \frac{\kappa^2}{\omega^2} - g_{\varphi\varphi} \bigg)
\bigg( g_{\varphi\varphi} - \frac{v^2 \kappa^2}{\omega^2} \bigg) \,.
\end{align}
In terms of $r=\cos \theta$ coordinate ($0 \leq r \leq 1$),
\begin{align}
r^{\prime 2} =&
\frac{\omega^2}{(1-v^2)^2}
\frac{[1+\vk_-^2(1-r^2)+\vk_+^2 r^2]^2}{1+\vk_-^2(1-r^2)}
\bigg( \frac{\kappa^2}{\omega^2} - g_{\varphi\varphi} \bigg)
\bigg( g_{\varphi\varphi} - \frac{v^2\kappa^2}{\omega^2} \bigg)
\,,
\end{align}
where
\begin{align}
  g_{\varphi\varphi} =& \frac{(1-r^2) \big( 1+\vk_-^2(1-r^2) \big)}{1+\vk_-^2(1-r^2)+\vk_+^2 r^2} \,.
% \qquad
%   g_{rr} = \frac{1}{(1-r^2)(1+\vk_-^2(1-r^2)+\vk_+^2 r^2)} \,.
\end{align}
Note that $g_{\varphi\varphi}$ is a monotonically decreasing function of $r$
from $g_{\varphi\varphi}=1$ ($r=0$) to $g_{\varphi\varphi}=0$ ($r=1$) under the condition $\vk_+ \geq \vk_-$.
In order to have a physical solution, $r'$ has to be real.
Namely $r^{\prime 2} \geq 0$.
This leads to the condition,
\begin{align}
  \frac{v^2\kappa^2}{\omega^2} \leq g_{\varphi\varphi}
\leq \frac{\kappa^2}{\omega^2} \,.
\end{align}
Thus, $v^2 \leq 1$ and $v^2 \leq \omega^2/\kappa^2$ are required.
We require that there exists two turning points;
this implies $\omega \geq \kappa$.
Two roots are given by
%$0 \leq r_\text{min} \leq r_\text{max} \leq 1$ such that
\begin{align}
  g_{\varphi\varphi} (r_\text{min}) = \frac{\kappa^2}{\omega^2}
\,,
\qquad
  g_{\varphi\varphi} (r_\text{max}) = \frac{v^2\kappa^2}{\omega^2}
\,.
\end{align}
A general root for the condition $g_{\varphi\varphi} (r_0) = {\cal C}$
is given by
\begin{align}
  r_0^2=&
\frac{1}{2\vk_-^2}
\bigg[
2\vk_-^2+1+{\cal C}(\vk_+^2- \vk_-^2)
- \sqrt{\big(2\vk_-^2+1+{\cal C}(\vk_+^2- \vk_-^2) \big)^2
-4\vk_-^2 (1+\vk_-^2)(1-{\cal C})}
\bigg] \,,
\end{align}
where we have taken the $-$ branch solution since
it has a smooth $\vk_- \rightarrow 0$ limit,
  $r_0^2 
= (1-{\cal C})/(1+\vk_+^2 {\cal C})$,
which agrees with the expression in \cite{Khouchen:2014kaa}.
Thus,
\begin{align}
  r_\text{min} =&
\sqrt{\frac{2\vk_-^2+1+{\frac{\kappa^2}{\omega^2}}(\vk_+^2- \vk_-^2)}{2\vk_-^2}
\bigg[1-\sqrt{1-\frac{4\vk_-^2 (1+\vk_-^2)(1-{\frac{\kappa^2}{\omega^2}})}{[2\vk_-^2+1+{\frac{\kappa^2}{\omega^2}}(\vk_+^2- \vk_-^2)]^2}}\bigg]}
\,,\\
  r_\text{max} =&
\sqrt{\frac{2\vk_-^2+1+{\frac{v^2\kappa^2}{\omega^2}}(\vk_+^2- \vk_-^2)}{2\vk_-^2}
\bigg[1-\sqrt{1-\frac{4\vk_-^2 (1+\vk_-^2)(1-{\frac{v^2\kappa^2}{\omega^2}})}{[2\vk_-^2+1+{\frac{v^2\kappa^2}{\omega^2}}(\vk_+^2- \vk_-^2)]^2}}\bigg]}
\,.
\label{r_max}
\end{align}

Now we calculate the various conserved charges.
The angular momentum $J(=J_1)$ is %($dy=\frac{dr}{|r'|}$)
\begin{align}
  J=& 2\hat{T} \int_{r_\text{min}}^{r_\text{max}} \frac{dr}{|r'|} \,
g_{\varphi\varphi} \omega (1- v h')
% \nn\\=&
% 2\hat{T} \int_{r_\text{min}}^{r_\text{max}} dr \,
% \sqrt{g_{rr} g_{\varphi\varphi}}
% \sqrt{\frac{g_{\varphi\varphi} - \frac{v^2\kappa^2}{\omega^2}}{\frac{\kappa^2}{\omega^2} - g_{\varphi\varphi}}}
\nn\\=&
2\hat{T} \int_{r_\text{min}}^{r_\text{max}} dr \,
\sqrt{g_{rr}(r) g_{\varphi\varphi}(r)}
\sqrt{\frac{g_{\varphi\varphi}(r) - g_{\varphi\varphi}(r_\text{max})}{g_{\varphi\varphi}(r_\text{min})-g_{\varphi\varphi}(r)}}
\,.
\end{align}
We also have
\begin{align}
  2\pi =& \int_{-\pi}^\pi d\sigma
= \int_{-\pi}^\pi dy
%\nn\\=&
=
2\int_{r_\text{min}}^{r_\text{max}} \frac{dr}{|r'|}
\nn\\=&
2 \frac{1-v^2}{\omega}
\int_{r_\text{min}}^{r_\text{max}} dr \,
\sqrt{\frac{g_{rr}}{\big( g_{\varphi\varphi}(r) - g_{\varphi\varphi}(r_\text{max})\big) \big( g_{\varphi\varphi}(r_\text{min})-g_{\varphi\varphi}(r)\big) }}
\,.
\end{align}
For later convenience, we define
\begin{align}
  g_{\varphi\varphi}(r) - g_{\varphi\varphi}(r_0) =&
% \frac{r^2-r_0^2}{(1+\vk_-^2(1-r^2)+\vk_+^2 r^2)(1+\vk_-^2(1-r_0^2)+\vk_+^2 r_0^2)}
% \nn\\& \hskip1em \times
% \bigg[
% \vk_-^2 \big(1+\vk_-^2+(\vk_+^2-\vk_-^2 )r_0^2 \big) r^2
% - (1+\vk_-^2)(1+\vk_+^2+\vk_-^2-\vk_-^2r_0^2)
% \bigg]
% \nn\\=&
\frac{r_0^2-r^2}{u_1(r^2) u_1(r_0^2)} u_3(r_0^2;r^2)
 \,,\\
  u_1(r^2)=&
1+\vk_+^2r^2 +\vk_-^2( 1-r^2) \,,
\qquad
  u_2(r^2)=
1+\vk_+^2 +\vk_-^2( 1-r^2) \,, \\
u_3(r_0^2;r^2) =&
(1+\vk_-^2) u_2(r_0^2) - \vk_-^2 u_1(r_0^2) r^2 \,.
\end{align}
% The factorization is not that complete as one parameter deformation case,
% $\vk_-=0$, where
% \begin{align}
%   g_{\varphi\varphi}(r) - g_{\varphi\varphi}(r_0) =&
% \frac{-(1+\vk_+^2)(r^2-r_0^2)}{(1+\vk_+^2 r^2)(1+\vk_+^2 r_0^2)} \,.
% \end{align}

The energy is given by (recall that $g_{tt}(\rho=0)=-1$)
\begin{align}
  E=& -P_t = \hat{T} \int_{-\pi}^\pi d\sigma \,
(-g_{tt}) \partial_\tau t
= -2\pi \kappa \hat{T} \,.
\end{align}
Namely, the constant $\kappa$ is related
to the energy as
\begin{align}
  \kappa=& - \frac{E}{2\pi \hat{T}} \,.
\end{align}

\paragraph{Infinite $J$ limit}

Following \cite{Khouchen:2014kaa},
we consider the infinite $J$ giant magnon.
This corresponds to $r_\text{min}=0$, namely,
\begin{align}
  \omega^2 = \kappa^2 = \frac{E^2}{(2\pi \hat{T})^2} \,. 
\end{align}
In this case, the spin $J$ and the constant factor $2\pi$ are
\begin{align}
&  J=
%   \frac{2\hat{T}}{\sqrt{1+\vk_-^2(1-r_\text{max}^2)+\vk_+^2 r_\text{max}^2}} 
% \nn\\& \times
% \int_{0}^{r_\text{max}} dr \,
% \frac{\sqrt{1+\vk_-^2(1-r^2)}}{1+\vk_-^2(1-r^2)+\vk_+^2 r^2}
% %\nn\\& \hskip2em \times
% \sqrt{\frac{r_\text{max}^2-r^2}{r^2}\frac{\vk_-^2 \big(1+\vk_-^2+(\vk_+^2-\vk_-^2 )r_\text{max}^2 \big) r^2
% - (1+\vk_-^2)(1+\vk_+^2+\vk_-^2-\vk_-^2r_\text{max}^2)
% }{\vk_-^2(r^2-1)-1-\vk_+^2}}
% \nn\\=&
  \frac{2\hat{T}}{\sqrt{u_1(r_\text{max}^2)}} 
%\nn\\& \times
\int_{0}^{r_\text{max}} dr \,
\frac{1}{u_1(r^2)}
%\nn\\& \hskip2em \times
\sqrt{\frac{r_\text{max}^2-r^2}{r^2}\frac{u_3(r_\text{max}^2; r^2)u_4(r^2)}{u_2(r^2)}}
\,.
\end{align}
\begin{align}
&  2\pi =
% 2 \frac{1-v^2}{\omega}
% \sqrt{1+\vk_-^2(1-r_\text{max}^2)+\vk_+^2 r_\text{max}^2}
% \nn\\& \times
% \int_{0}^{r_\text{max}} dr \,
% \sqrt{\frac{1+\vk_-^2(1-r^2)+\vk_+^2 r^2}{r^2(1-r^2)(r_\text{max}^2-r^2)}
% }
% \nn\\& \times
% \sqrt{\frac{1}{\big[\vk_-^2 \big(1+\vk_-^2+(\vk_+^2-\vk_-^2 )r_\text{max}^2 \big) r^2
% - (1+\vk_-^2)(1+\vk_+^2+\vk_-^2-\vk_-^2r_\text{max}^2) \big] \big(\vk_-^2(r^2-1)-1-\vk_+^2 \big)}}
% \nn\\=&
2 \frac{1-v^2}{\omega}
\sqrt{u_1(r_\text{max}^2)}
%\nn\\& \times
\int_{0}^{r_\text{max}} dr \,
\sqrt{\frac{1}{r^2(r_\text{max}^2-r^2)}
\frac{u_4(r^2)}{u_3(r_\text{max}^2;r^2)u_2(r^2)}}
\,,
\end{align}
where both of them are
divergent quantities.
We need to choose a constant $K$
such that $2\pi K - J$ becomes finite.
This $K$ will be related to the energy $E$
and it will give a dispersion relation
$E-J=\text{finite}$.
The divergence is due to the lower end of the integral $r=0$,
and in order to cancel this divergence $K$ is chosen to be
\begin{align}
  K= \omega \hat{T} \,.
\end{align}
This is the same result as in \cite{Khouchen:2014kaa}
except the overall sign.
(In \cite{Khouchen:2014kaa}, their $J$ has an opposite sign.
But we may choose the sign of $K$ for $J$ to have a positive value.)
We choose the positive root of $\omega=E/(2\pi \hat{T})$,
and
\begin{align}
&  E-J
%  =&
%   \frac{2\hat{T}}{\sqrt{u_1(r_\text{max}^2)}} 
% \int_{0}^{r_\text{max}} dr \,
% \bigg(
% r_\text{max}^2u_2(r_\text{max}^2) f_\text{int}(r)
%  -  J_\text{int}(r)
% \bigg)
\nn\\=&
2\hat{T} 
\int_{0}^{r_\text{max}} dr \,
\frac{1}{r \sqrt{u_1(r_\text{max}^2)}} \sqrt{\frac{1+\vk_-^2(1-r^2)}{(r_\text{max}^2-r^2)u_2(r^2) u_3(r_\text{max}^2;r^2)}}
%\nn\\& \hskip2em \times
\bigg[
r_\text{max}^2u_2(r_\text{max}^2) 
-\frac{(r_\text{max}^2-r^2) u_3(r_\text{max}^2;r^2)}{u_1(r^2)}
\bigg]
 \,.
\end{align}
We may want to represent the dispersion relation $E-J$ as
 a function of the momentum (or the deficit angle of the string configuration), but
the current expression is too complicated to analyze analytically.
We thus consider perturbative corrections with respect to $\vk_-$.
First,
\begin{align}
  r_\text{max}^2 =&
% \frac{1-v^2}{1+v^2 \vk_+^2}
% \bigg[
% 1+ \frac{v^4 \vk_+^2 (1+\vk_+^2)}{(1+v^2 \vk_+^2)^2} \vk_-^2
% +\mathcal{O}(\vk_-^4)
% \bigg]=
r_\text{max,0}^2
\bigg[
1+ \frac{v^4 \vk_+^2 (1+\vk_+^2)}{(1+v^2 \vk_+^2)^2} \vk_-^2
-\frac{v^6 \vk_+^2 (1+\vk_+^2)\big( 1+(2-v^2)\vk_+^2 \big)}{(1+v^2 \vk_+^2)^4} \vk_-^4
+\mathcal{O}(\vk_-^6)
\bigg]
\,,
\end{align}
where $r_\text{max,0}^2 = \frac{1-v^2}{1+v^2 \vk_+^2}$.
In order to avoid a unnecessary divergence, 
we first expand the integrands and integrate them to $\rmax$,
and then expand the results in terms of $\vk_-$.
% \footnote{%
% If the first expand $\rmax$ in the definite integral, we encounter
% unnecessarily divergent terms.
% For example, consider
% \begin{align}
%   \int_0^X dx \frac{x}{\sqrt{X^2-x^2}} = X \,.
% \end{align}
% When we expand $X=X_0+\epsilon$,
% we have
% \begin{align}
%   \int_0^{X_0+\epsilon}dx  \bigg[ \frac{x}{\sqrt{X_0^2-x^2}} - \frac{xX_0}{(X_0^2-x^2)^{3/2}}+\mathcal{O}(\epsilon^2) \bigg]
% = X_0
% + \epsilon \lim_{x \rightarrow X_0{}^-} \frac{x}{\sqrt{X_0^2-x^2}}
% + \epsilon \bigg[-\frac{X_0}{\sqrt{X_0^2-x^2}} \bigg]^{X_0}_0
% = X_0+\epsilon \,.
% \end{align}
% In the middle, the second term and the upper limit of the third term are divergent,
% but cancel out.
% It is better to substitute $X=X_0+\epsilon$ in the original
%  expression.}
This leads to
\begin{align}
E-J
% =& 2\hat{T} \bigg[
% \frac{1}{\vk_+} \tanh^{-1}\ \bigg( \frac{\vk_+ \rmax}{\sqrt{1+\vk_+^2 \rmax^2}}  \bigg)
% \nn\\& \hskip3em
% +\frac{\vk_-^2}{\vk_+^3}
% \tanh^{-1}\ \bigg( \frac{\vk_+ \rmax}{\sqrt{1+\vk_+^2 \rmax^2}}  \bigg)
% \nn\\& \hskip3em
% -\frac{\vk_-^2 \rmax}{2\vk_+^2(1+\vk_+^2)\sqrt{1+\vk_+^2\rmax^2}}
% \big(2+3\vk_+^2+\vk_+^4(1+\rmax^2-\rmax^4) \big)
% \nn\\& \hskip3em
% +\mathcal{O}(\vk_-^4)
% \bigg]
=&
2\hat{T} \bigg[
\frac{\vk_+^4+\vk_+^2\vk_-^2+\vk_-^4}{\vk_+^5} \sinh^{-1}\ \big( \vk_+ \rmaxz  \big)
\nn\\&
-\frac{\vk_-^2  \rmaxz \sqrt{1+\vk_+^2\rmaxz^2}}{2\vk_+^2 (1+\vk_+^2) }
\big(2+3\,{\vk_+}^{2}-2 \vk_+^2 \rmaxz^2 \big)
\nn\\&
+\frac{\vk_-^4  \rmaxz \sqrt{1+\vk_+^2\rmaxz^2}}{24\vk_+^4 (1+\vk_+^2)^2 }
\big[2 \vk_+^6 \rmaxz^2
   \left(24 \rmaxz^4-56 \rmaxz^2+35\right)
\nn\\& \hskip6em
+\vk_+^4 \left(-8 \rmaxz^4+20
   \rmaxz^2-15\right)
+16 \vk_+^2 \left(\rmaxz^2-3\right)-24
\big]
\nn\\&
+\mathcal{O}(\vk_-^6)
\bigg]
\,,
\end{align}
where we have used $v^2 = (1-\rmaxz^2)/(1+\vk_+^2\rmaxz^2)$.

The momentum corresponds to the angle spanned by asymptotic directions
of the string,
\begin{align}
  p= & \Delta \varphi
= \int d\varphi
=2\int_{r_\text{min}}^{r_\text{max}} \frac{dr}{|r'|} \varphi'
% \nn\\=&
% 2v \int_{0}^{r_\text{max}} dr \,
% \frac{\sqrt{1+\vk_-^2(1-r^2)}}{1+\vk_-^2(1-r^2)+\vk_+^2r^2}
% \frac{1}{g_{\varphi\varphi}}\sqrt{\frac{1-g_{\varphi\varphi}}{g_{\varphi\varphi}-v^2}}
% \nn\\=&
% 2v\sqrt{1+\vk_+^2 r_\text{max}^2+\vk_-^2(1-r_\text{max}^2)}
% \nn\\& \times  \int_{0}^{r_\text{max}} dr \,
%        \frac{r}{1-r^2}\sqrt{ \frac{1+\vk_+^2+\vk_-^2(1-r^2)}{\big(r_\text{max}^2-r^2\big)\big( 1+\vk_-^2(1-r^2)\big)}}
% \nn\\& \hskip2em \times
% \frac{1}{\sqrt{ (1+\vk_-^2) \big(1+\vk_+^2+\vk_-^2\big(1-r_\text{max}^2\big)\big)
% - \vk_-^2\big(1+\vk_+^2r_\text{max}^2+\vk_-^2\big(1-r_\text{max}^2\big) \big)  r^2 }}
\nn\\=&
2v 
 \int_{0}^{r_\text{max}} dr \,
     \sqrt{u_1(r_\text{max}^2)}  \frac{r}{1-r^2}
\sqrt{\frac{u_2(r^2)}{\big(r_\text{max}^2-r^2\big) u_3(r_\text{max}^2;r^2) u_4(r^2)}}
\,.
\end{align}
In the $\vk_- \rightarrow 0$ limit, it is reduced to
$r_\text{max,0} = \sin \frac{p}{2}$ \cite{Khouchen:2014kaa}.
% This leading order relation does not involve $\vk_+$, and is the same as
% the undeformed one.
% However, in the case of two parameter deformation, 
% the relation becomes much more complicated.
Small $\vk_-$ corrections for $p$ are calculated as
\begin{align}
  p=&
2\arcsin \rmaxz
\nn\\&
-\frac{\vk_-^2 \rmaxz
   \sqrt{1-\rmaxz^2} \left(2 \vk_+^2 \rmaxz^2+1\right)}{\vk_+^2+1}
\nn\\&
-\frac{\vk_-^4 \rmaxz \sqrt{1-\rmaxz^2} \left(48 \vk_+^4 \rmaxz^6-8
   \vk_+^2 \left(7 \vk_+^2-6\right) \rmaxz^4+\left(6-60 \vk_+^2\right)
   \rmaxz^2-9\right)}{12 \left(\vk_+^2+1\right)^2}
\nn\\&
+\mathcal{O}(\vk_-^6) \,.
\end{align}
This relation can be inverted to $\rmaxz = \sin \frac{p}{2} + \cdots$.
By substituting this into $E-J$ result, we obtain perturbative corrections
to the dispersion relation,
\begin{align}
&  E-J
\nn\\=&
2\hat{T} \bigg[
\frac{\vk_+^4+\vk_+^2\vk_-^2+\vk_-^4}{\vk_+^5} \sinh^{-1}\ \bigg( \vk_+ \sin \left(\frac{p}{2}\right)  \bigg)
\nn\\&
+\frac{\vk_-^2 \sin \left(\frac{p}{2}\right) }{16 \vk_+^2 \left(\vk_+^2+1\right)
   \sqrt{\vk_+^2 \sin ^2\left(\frac{p}{2}\right)+1}}
\bigg[
\vk_+^4 \sin
   \left(\frac{p}{2}\right) \left(15 \sin ^3\left(\frac{p}{2}\right)-10 \sin
   ^2\left(\frac{p}{2}\right)-20 \sin \left(\frac{p}{2}\right)+7\right)
\nn\\& \hskip4em
-2 \vk_+^2
   \left(\sin ^2\left(\frac{p}{2}\right)+3 \sin
   \left(\frac{p}{2}\right)+8\right)-16
\bigg]
\nn\\&
+\frac{\vk_-^4 \sin \left(\frac{p}{2}\right)}{1536 \vk_+^4
   \left(\vk_+^2+1\right)^2 \left(\vk_+^2 \sin
   ^2\left(\frac{p}{2}\right)+1\right)^{3/2}}
\nn\\& \times
 \bigg[\vk_+^{10} \sin   ^3\left(\frac{p}{2}\right)
 \bigg(2700 \sin ^7\left(\frac{p}{2}\right)-3360 \sin
   ^6\left(\frac{p}{2}\right)-4211 \sin ^5\left(\frac{p}{2}\right)+4569 \sin
   ^4\left(\frac{p}{2}\right)
\nn\\& \hskip8em
+2843 \sin ^3\left(\frac{p}{2}\right)-1941 \sin
   ^2\left(\frac{p}{2}\right)-147 \sin \left(\frac{p}{2}\right)-69\bigg)
\nn\\& \hskip2em
+\vk_+^8 \sin
   \left(\frac{p}{2}\right) 
\bigg(4423 \sin ^7\left(\frac{p}{2}\right)-5964 \sin
   ^6\left(\frac{p}{2}\right)-5347 \sin ^5\left(\frac{p}{2}\right)+7167 \sin
   ^4\left(\frac{p}{2}\right)
\nn\\& \hskip8em
+3363 \sin ^3\left(\frac{p}{2}\right)-2997 \sin
   ^2\left(\frac{p}{2}\right)-69\bigg)
\nn\\& \hskip2em
+4 \vk_+^6 \sin \left(\frac{p}{2}\right)
   \bigg(519 \sin ^5\left(\frac{p}{2}\right)-699 \sin ^4\left(\frac{p}{2}\right)-546 \sin
   ^3\left(\frac{p}{2}\right)+660 \sin ^2\left(\frac{p}{2}\right)
\nn\\& \hskip8em
-224 \sin
   \left(\frac{p}{2}\right)-222\bigg)
\nn\\& \hskip2em
-4 \vk_+^4 \left(39 \sin
   ^4\left(\frac{p}{2}\right)+54 \sin ^3\left(\frac{p}{2}\right)+1024 \sin
   ^2\left(\frac{p}{2}\right)-45 \sin \left(\frac{p}{2}\right)+384\right)
\nn\\& \hskip2em
-1024 \vk_+^2
   \left(2 \sin ^2\left(\frac{p}{2}\right)+3\right)-1536\bigg)
\bigg]
\nn\\&
+\mathcal{O}(\vk_-^6) \,.
\end{align}
The first term is a generalization of the result of \cite{Khouchen:2014kaa};
the coefficient has some $\vk_-$ corrections.
The other terms are corrections in terms of $\sin \left(\frac{p}{2}\right)$.
It is interesting to see how this complicated dispersion relation can be obtained
via a dual gauge theory, but at this moment it is not clear.

\subsection{Two-spin ``spiky'' string solution}
\label{sec:two-spin-spily}

We next try the two spin ansatz of \cite{Dai:2014twa},
\begin{align}
  t=& \tau+h_1(y) \,,
\quad \rho=\rho(y) \,,
\quad
\psi=\omega [ \tau+h_2(y)] \,,
\quad
\phi= \Omega \tau \,,
\quad
\theta=\frac{\pi}{2} \,,
\end{align}
where $y=\sigma-v\tau$.
In this case, as we will see, we find several types of hanging string solutions,
rather than a spiky string solution.
The action is
\begin{align}
 I =&
  -\frac{\hat{T}}{2} \int d\tau dy \bigg[
g_{tt} \big( (1-v^2) h_1^{\prime 2} +2v h_1' -1 \big)
+g_{\rho\rho} (1-v^2)\rho^{\prime 2}
+g_{\psi\psi} \omega^2 \big((1-v^2) h_2^{\prime 2} +2v h_2'-1 \big)
\nn\\& \hskip4em
+2g_{t\psi}
\omega  \big((1-v^2)h_1'h_2' +v( h_1' + h_2')-1 \big)
-\Omega^2
\bigg] \,.
% \nn\\=&  -\frac{\hat{T}}{2} \int d\tau dy \bigg[
% \frac{1}{1+\vk_-^2\cosh^2\rho -\vk_+^2\sinh^2 \rho}
% \bigg(
% -\cosh^2\rho \big( 1+\vk_-^2\cosh^2\rho \big)
% \big( (1-v^2) h_1^{\prime 2} +2v h_1' -1 \big)
% \nn\\& \hskip10em
% +(1-v^2)\rho^{\prime 2}
% +\omega^2 \sinh^2\rho(1-\vk_+^2\sinh^2\rho )
%  \big((1-v^2) h_2^{\prime 2} +2v h_2'-1 \big)
% \nn\\& \hskip10em
% +2\varkappa_+\varkappa_-\sinh^2\rho \cosh^2\rho
% \cdot \omega
%  \big((1-v^2)h_1'h_2' +v(h_1' + h_2')-1 \big)
% \bigg)
% %\nn\\& \hskip 7em
% -\Omega^2
% \bigg] \,.
\end{align}
The equations of motion for $h_1$ and $h_2$, after once integrated, are
% \begin{align}
% & \frac{d}{dy} \bigg[
% g_{tt} \big((1-v^2)h_1' +v \big)
%  +\omega g_{t\psi} \big((1-v^2) h_2' + v \big)
% \bigg]=0 
% \label{eq:h1_1}
% \,,\\
% & \frac{d}{dy}
% \bigg[
% \omega^2 g_{\psi\psi} \big((1-v^2)h_2' +v \big)
%  +\omega g_{t\psi} \big((1-v^2) h_1' + v \big)
% \bigg]=0 \,.
% \label{eq:h2_1}
% \end{align}
% They are integrated once,
\begin{align}
&
g_{tt}  \big((1-v^2)h_1' +v \big)
 +\omega g_{t\psi} \big((1-v^2) h_2' + v \big)
=-c_1
\label{eq:h1_2}
 \,,  \\ &
\omega^2 g_{\psi\psi} \big((1-v^2)h_2' +v \big)
 +\omega g_{t\psi} \big((1-v^2) h_1' + v \big)
= \omega^2 c_2 \,,
\label{eq:h2_2}
\end{align}
where $c_1$ and $c_2$ are constants of integration.
The negative sign for $c_1$ and
an extra $\omega^2$ for $c_2$ are for later convenience.
Thus,
\begin{align}
h_1'=&
\frac{1}{1-v^2} \bigg[
\frac{1-\vk_+^2 \sinh^2\rho}{\cosh^2\rho}c_1 
+\omega \vk_+ \vk_- c_2
-v
\bigg] \,,\\
  h_2'=&
\frac{1}{1-v^2} \bigg[
-\frac{\vk_+ \vk_- c_1}{\omega}
+ \frac{1+\vk_-^2\cosh ^2 \rho}{ \sinh^2 \rho} c_2
- v
\bigg] \,.
\end{align}

By eliminating $\rho^{\prime 2}$ from two Virasoro constraints,
\begin{align}
  0=&
g_{tt} \big( (1+v^2) h_1^{\prime 2} - 2v h_1' +1 \big)
+g_{\rho\rho} (1+v^2) \rho^{\prime 2}
+ g_{\psi\psi} \omega^2 \big( (1+v^2) h_2^{\prime 2} - 2v h_2' +1 \big)
\nn\\& \hskip2em
+ 2 g_{t\psi}\omega \big( (1+v^2) h_1'h_2' - v(h_1' + h_2') +1 \big)
+\Omega^2
\,, \\
0=&
g_{tt} h_1' \big( 1- v h_1^{\prime} \big) 
-g_{\rho\rho} v \rho^{\prime 2}
+ \omega^2 g_{\psi\psi} h_2' \big(1- v h_2' \big) 
+ \omega g_{t\psi} (h_1'+h_2' -2v h_1' h_2') \,,
\end{align}
we obtain
\begin{align}
g_{tt}  \big((1-v^2)h_1' +v \big)
+\omega^2 g_{\psi\psi} \big((1-v^2)h_2' +v \big)
+\omega g_{t\psi} \big((1-v^2)(h_1'+ h_2') + 2v \big)
+ v \Omega^2=0 \,.
\end{align}
Using \eqref{eq:h1_2} and \eqref{eq:h2_2}, we find
the relation for the constants of integration,
\begin{align}
  c_1 - \omega^2 c_2 = v \Omega^2  \,.
\end{align}
From now on, we set $\Omega=1$ for simplicity.
We take the solution $c_1=v$ and $c_2=0$ \cite{Ryang:2006yq} 
which satisfies 
the condition for forward propagation of the string,
\begin{align}
  \frac{dt}{d\tau} =& 
\frac{v}{1-v^2} \bigg[
\frac{1}{v} - \frac{1-\vk_+^2 \sinh^2\rho}{\cosh^2\rho} c_1
- \vk_+\vk_- \omega c_2 \bigg] > 0 \,.
\end{align}
The solutions are now
\begin{align}
h_1'=&
% \frac{v}{1-v^2} \bigg[
% \frac{1-\vk_+^2 \sinh^2\rho}{\cosh^2\rho}
% -1
% \bigg]
% =
 -(1+\vk_+^2) \frac{v}{1-v^2} \tanh^2 \rho
 \,,\qquad
  h_2'=
% \frac{v}{1-v^2} \bigg[
% -\frac{\vk_+ \vk_-}{\omega}
% - 1
% \bigg]
% =
-\bigg(1+\frac{\vk_+ \vk_-}{\omega} \bigg) \frac{v}{1-v^2}
 \,.  
\end{align}
$\rho'$ can be obtained from these two solution and the second Virasoro constraint,
\begin{align}
(\rho')^2  =&
% v^{-1} \big[ -\cosh^2\rho ( 1+ \vk_-^2 \cosh^2\rho) h_1' (1-v h_1')
% +\omega^2 \sinh^2\rho (1-\vk_+^2\sinh^2 \rho) h_2' (1-v h_2')
% \nn\\&
% +\omega \vk_+ \vk_- \sinh^2\rho \cosh^2\rho
% \big(h_1' + h_2'-2v h_1' h_2' \big) \big]
% \nn\\=&
\frac{\tanh^2\rho}{(1-v^2)^2}
f(\rho)
\,,\\
f(\rho)=&
(\omega \vk_+-\vk_-)^2 \cosh^4 \rho
+ (1+\vk_+^2) \big( 1-\omega^2 +v^2(\vk_+^2-\vk_-^2) \big) \cosh^2\rho
-v^2(1+\vk_+^2)^2 \,.  
\end{align}
Note that $\rho' \rightarrow \pm \infty$ for $\rho \rightarrow \infty$.
We may write
\begin{align}
  \rho' =& \pm \frac{\tanh \rho}{1-v^2} \sqrt{f(\rho)} \,,
\end{align}
and $f(\rho) =0$ has two roots as a function of $\cosh^2\rho$ as 
\begin{align}
  \cosh^2 \rho_\pm =&
\frac{1+\vk_+^2}{2(\omega \vk_+-\vk_-)^2 }
\bigg[-1 + \omega^2 -v^2 (\vk_+^2-\vk_-^2)
\nn\\& \hskip3em
\pm \sqrt{\big((1-\omega)^2 + v^2 (\vk_+-\vk_-)^2 \big) \big((1+\omega)^2 + v^2 (\vk_++\vk_-)^2 \big) }  \bigg] \,,
\end{align}
where $\omega \vk_+-\vk_- \neq 0$ is assumed.
%The case of $\omega \vk_+-\vk_- =0$ is considered in Appendix \ref{sec:special-case-1}.

Let us examine the allowed region of $\rho$.
The original metric has a singularity at
(recall that $\vk_+ \geq \vk_-$ is assumed)
\begin{align}
  \rho_s =& \sinh^{-1} \bigg[\sqrt{\frac{1+\vk_-^2}{\vk_+^2-\vk_-^2}} \bigg] \,,
% \qquad
% \bigg( \cosh^2 \rho_s = \frac{1+\vk_+^2}{\vk_+^2-\vk_-^2} \bigg)
\end{align}
It is not difficult to check that $\rho_\pm$ and $\rho_s$ satisfy the following inequalities,
\begin{align}
\cosh^2  \rho_- \leq 0 \,,
\qquad
\rho_s \geq \rho_+ \,.
\label{eq:rho_ineq}
\end{align}
Therefore, $\rho_-$ is not real.
Since $(\rho')^2 \rightarrow \infty$ for $\rho \rightarrow \infty$,
$\rho'$ is real for $\rho_+ \leq \rho$.
Thus, the possible classical solutions exist in the following regions:
\begin{align}
  \text{(I)}: \quad \rho_+ \leq \rho \leq \rho_s
\qquad \qquad
\text{(II)}: \quad \rho_s \leq \rho < \infty 
\end{align}
where the region (I) is valid if $\rho_+$ is real.
This condition for $\rho_+$ to be real is
\begin{align}
\omega \geq -\vk_+ \vk_- + \sqrt{(1-v^2)(1+\vk_+^2)(1+\vk_-^2)} \,,
\qquad
\omega \leq -\vk_+ \vk_- - \sqrt{(1-v^2)(1+\vk_+^2)(1+\vk_-^2)} \,.
\label{eq:reality_cond_omega}
\end{align}
Otherwise, there exists no turning point, and we consider
\begin{align}
  \text{(I)}': \quad 0 \leq \rho \leq \rho_s \,.
\end{align}
This can also be viewed as a hanging string solution from the singular surface to the center of AdS.
% ,
% or a spiky string solution that reachres to the location of the singular surface.

\begin{figure}[tbh]
  \centering
  \includegraphics[scale=0.7]{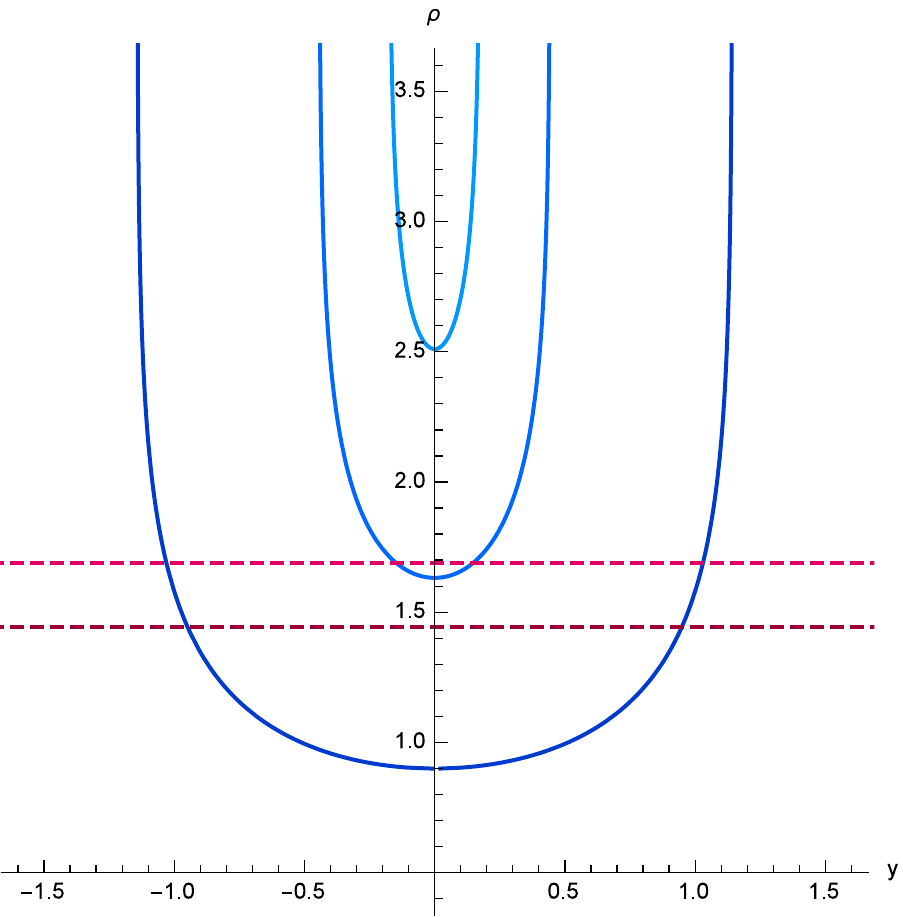}
 \put(-172,140){\makebox(0,0){\tiny $\vk_-=0$}}
 \put(-135,160){\makebox(0,0){\tiny $\vk_-=0.3$}}
 \put(-115,180){\makebox(0,0){\tiny $\vk_-=0.5$}}
 \put(-20,75){\makebox(0,0){\tiny $\vk_-=0.3$}}
 \put(-20,52){\makebox(0,0){\tiny $\vk_-=0$}}
\hspace{2em}
  \includegraphics[scale=0.7]{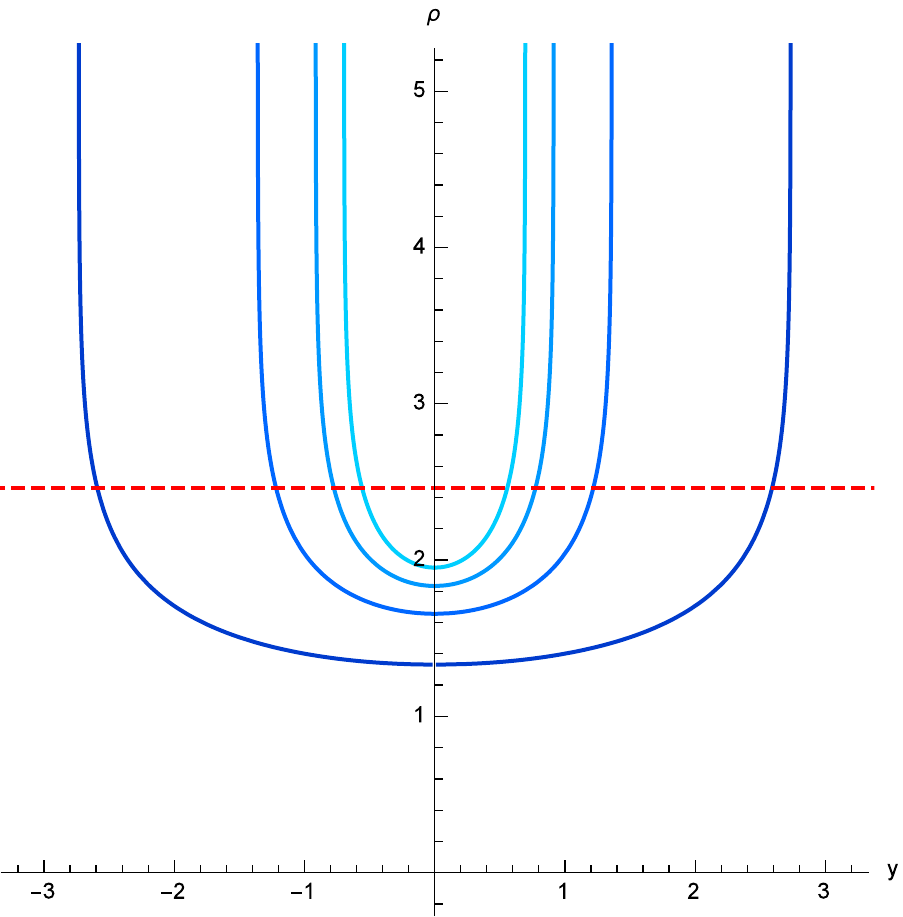}
 \put(-182,140){\makebox(0,0){\tiny $\omega=1.02$}}
 \put(-147,160){\makebox(0,0){\tiny $\omega=1.04$}}
  \caption{Hanging string solutions with
$v=0.1$.
(Left)  $\omega=1.3$ and $\vk_+=0.5$, and 
varied values of $\vk_-=0,0.3,0.5$.
(from darker color to brighter).
The horizontal dashed lines indicate the locations of
the corresponding singular surfaces;
$\rho_s=1.44364$ ($\vk_-=0$), $1.68735$ ($\vk_-=0.3$), and $\infty$ ($\vk_-=0.5$).
(Right)
Varying $\omega=1.02, 1.04, 1.06, 1.08$ (from the outermost darkest one to inside)
with fixed $\vk_-=0.1$ and $\vk_+=0.2$.
The value of $\rho_s=2.45875$.}
  \label{fig:spikes_km_chg1}
\end{figure}

\begin{figure}[tbh]
  \centering
  \includegraphics[scale=0.7]{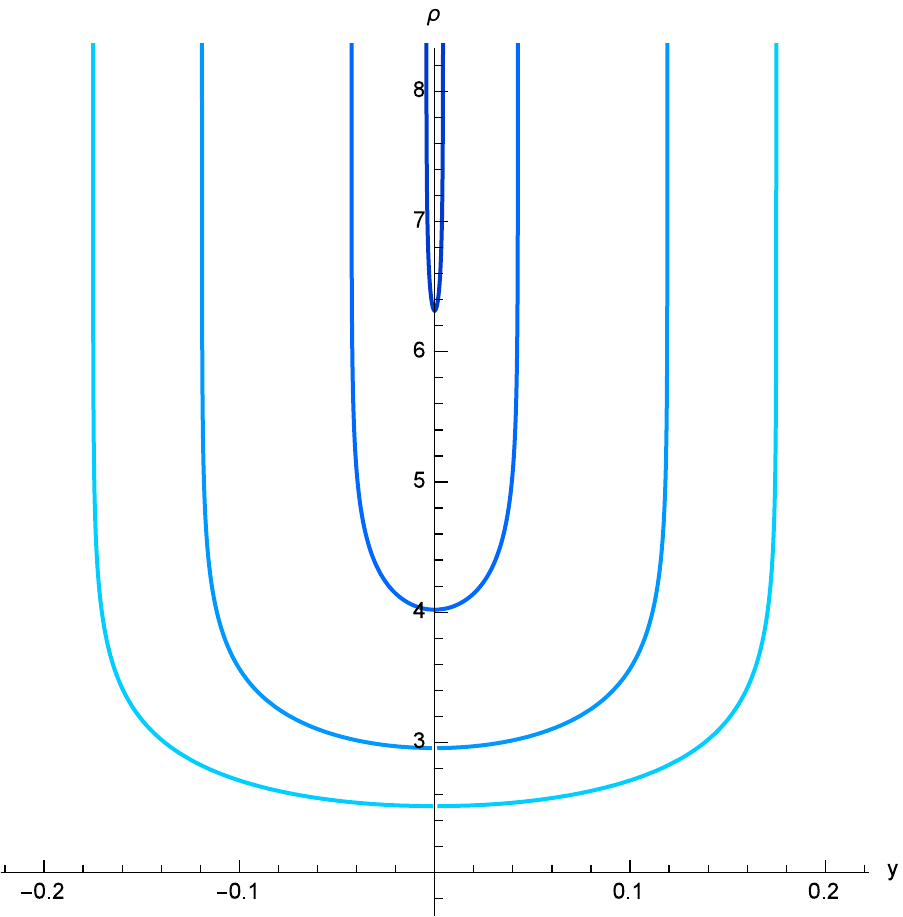}
 \put(-180,160){\makebox(0,0){\tiny $\vk_+=0.5$}}
 \put(-150,180){\makebox(0,0){\tiny $\vk_+=0.3$}}
% \put(-1,160){\makebox(0,0){\tiny $\vk_-=0.1$}}
\hspace{2em}
  \includegraphics[scale=0.7]{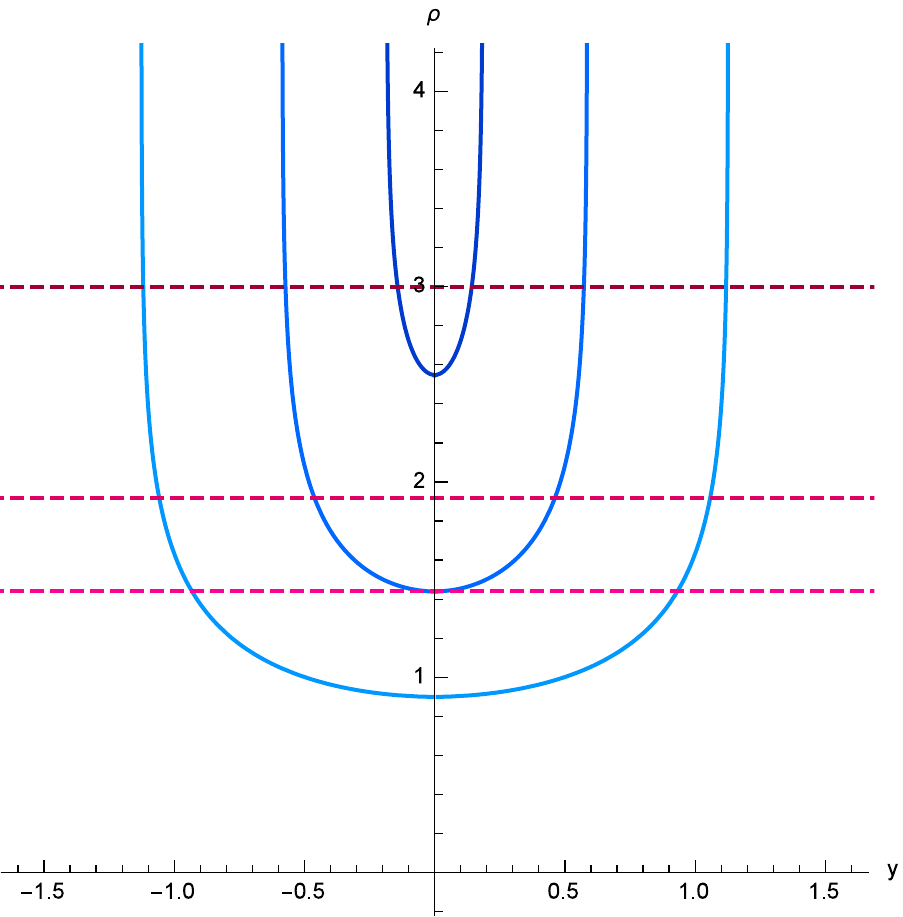}
 \put(-173,140){\makebox(0,0){\tiny $\vk_+=0.5$}}
 \put(-130,180){\makebox(0,0){\tiny $\vk_+=0.3$}}
 \put(0,133){\makebox(0,0){\tiny $\vk_+=0.1$}}
 \put(0,90){\makebox(0,0){\tiny $\vk_+=0.3$}}
 \put(0,60){\makebox(0,0){\tiny $\vk_+=0.5$}}
  \caption{Hanging string solutions in special cases.
(Left) $\vk_+=\vk_-$ case. Namely $\rho_s=\infty$.
The parameters are $v=0.1$, $\omega=1.3$, and $\vk_+=\vk_-=0.01,0.1,0.3,0.5$
(from the innermost darkest color to outside).
(Right) $\vk_-=0$ case; namely the one-parameter deformation case.
The parameters are $v=0.1$, $\omega=1.3$, and $\vk_+=0.1,0.3,0.5$
(from the innermost darkest color to outside).
The corresponding locations of the singular surface are
$\rho_s=2.99822, 1.9189, 1.44364$. }
  \label{fig:spikes_specials}
\end{figure}

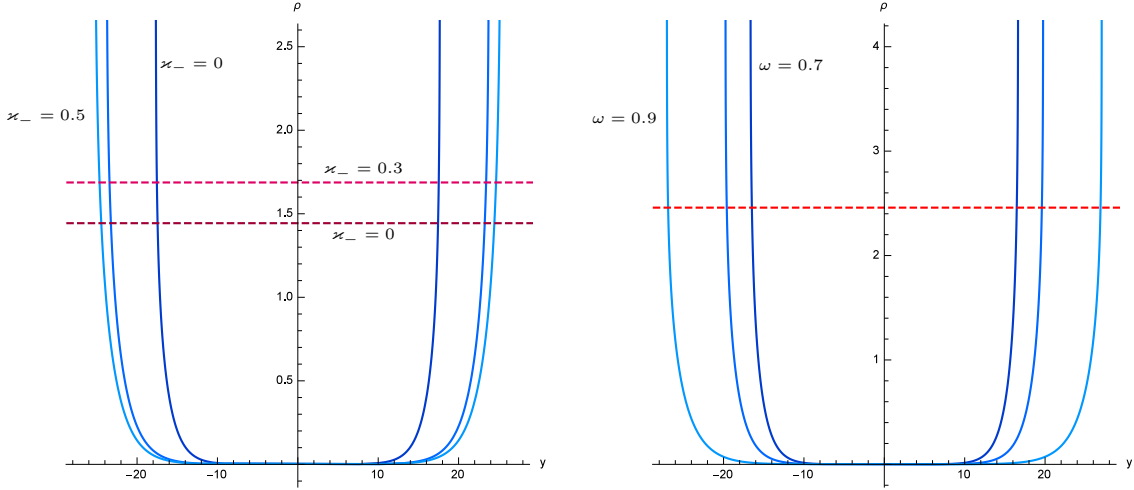
\begin{figure}[tbh]
  \centering
  \includegraphics[scale=0.7]{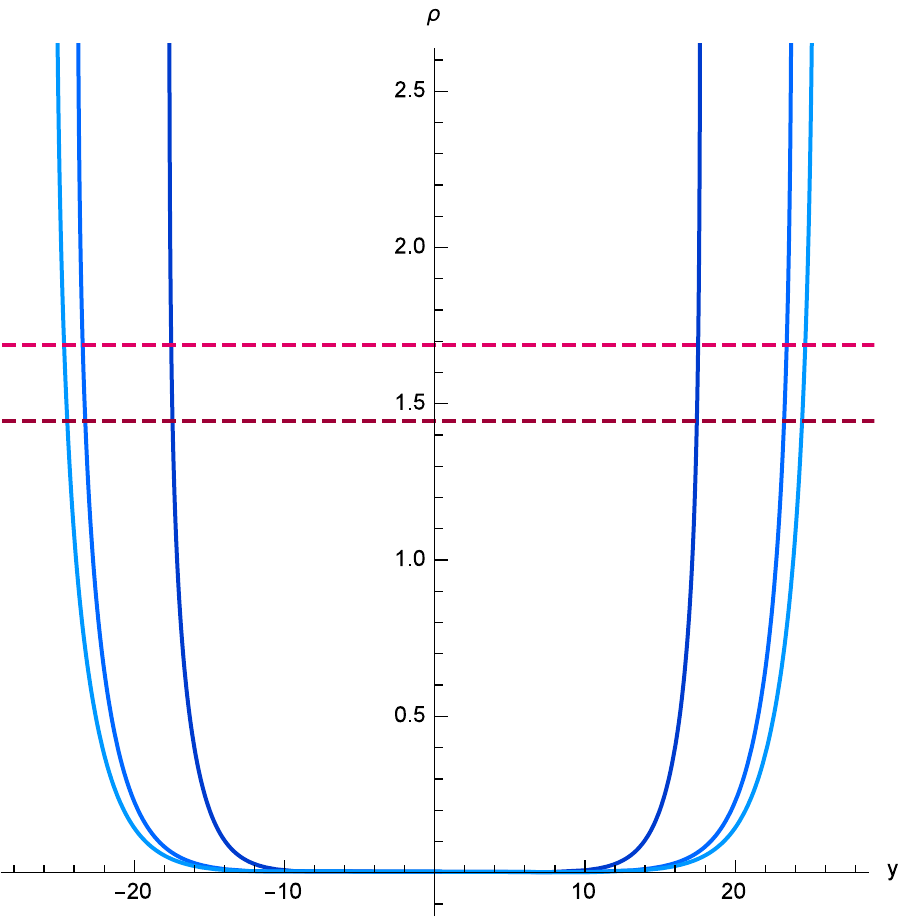}
 \put(-190,140){\makebox(0,0){\tiny $\vk_-=0.5$}}
 \put(-135,160){\makebox(0,0){\tiny $\vk_-=0$}}
 \put(-70,120){\makebox(0,0){\tiny $\vk_-=0.3$}}
 \put(-70,95){\makebox(0,0){\tiny $\vk_-=0$}}
\hspace{3em}
  \includegraphics[scale=0.7]{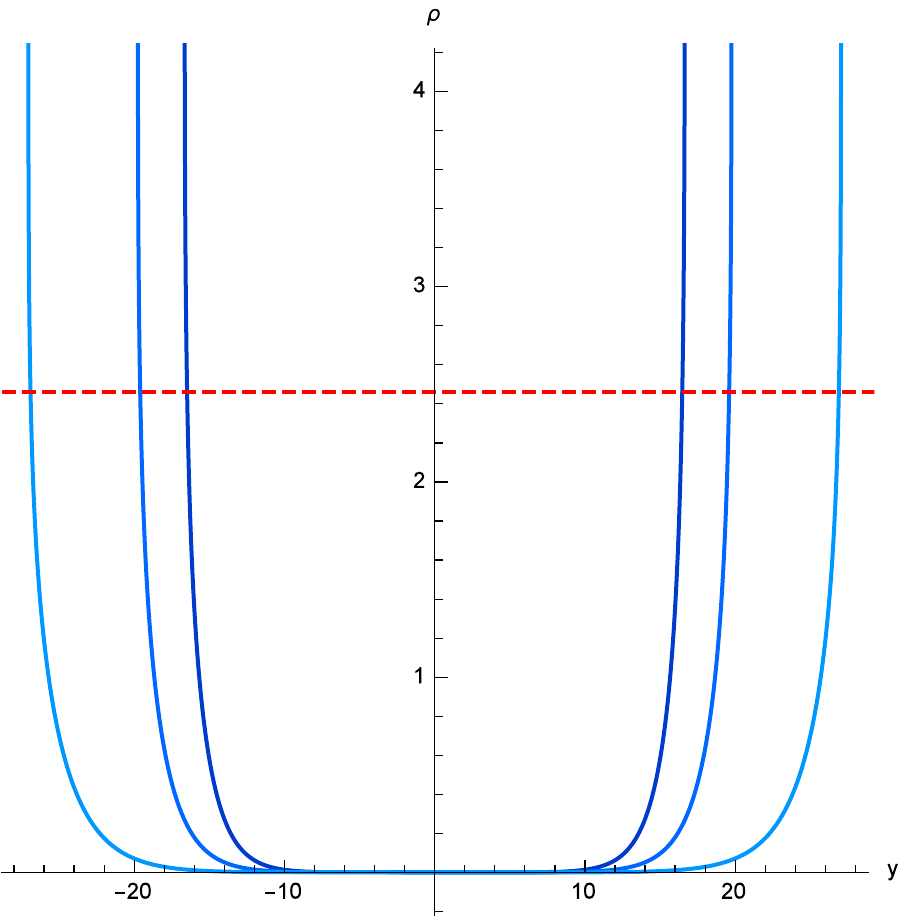}
 \put(-193,140){\makebox(0,0){\tiny $\omega=0.9$}}
 \put(-130,160){\makebox(0,0){\tiny $\omega=0.7$}}
  \caption{Hanging string solutions with the case (I)${}'$.
 In this case, the solutions reach the center of AdS, $\rho=0$.
% This figure is written as a hanging solution from $\rho_s$.
% This can be viewed as a spike solution towards the singular surface.
(Left) Varying $\vk_-$. 
$v=0.1$, $\omega=0.9$ and $\vk_+=0.5$, and 
varied values of $\vk_-=0,0.3,0.5$.
(from the innermost darkest one to outside).
The corresponding values of $\rho_s$ are
$\rho_s=1.44364, 1.68735, \infty$.
(Right) Varying $\omega$.
$v=0.1$, $\vk_-=0.1$, $\vk_+=0.2$ and
$\omega=0.7, 0.8, 0.9$ (from the innermost darkest one to outside).
The corresponding $\rho_s=2.45875$.}
  \label{fig:spikes_Ip}
\end{figure}

The results with real $\rho_+$ cases are summarized in Fig.~\ref{fig:spikes_km_chg1}
and \ref{fig:spikes_specials}.
It contains the both region (I) and (II), and they are separated by the locations of the singular surfaces (the horizontal dashed lines in the figures).
Fig.~\ref{fig:spikes_km_chg1} shows the profiles for a general parameter setting,
while Fig.~\ref{fig:spikes_specials} displays two special cases.
The left one is for $\vk_+=\vk_-$, and then the singular surface is pushed to infinity.
The right one is the case with $\vk_-=0$ which is reduced to the one-parameter deformation
result.
When $\omega$ is small $\rho_+$ becomes imaginary. So the solutions run from $\rho=0$
to $\rho_s$ (and further).
This case is summarized in Fig.~\ref{fig:spikes_Ip}.

\paragraph{Dispersion relation}
\label{sec:dispersion-relation}

The conserved charges, in the region (I), are
% \footnote{%
% For constant $\tau$, $d\sigma = dy$.
% Again, there may be minus sign for the definition of $E$.}
\begin{align}
  E=&
- \hat{T} \int d\sigma \, 
g_{tt} \dot{t}
% \nn\\=&
%  \hat{T} \int dy \, 
% \frac{\cosh^2\rho \big( 1+\vk_-^2\cosh^2\rho \big)}{1+\vk_-^2\cosh^2\rho -\vk_+^2\sinh^2 \rho}
% (1-v h_1')
%\nn\\=&
=
\frac{\hat{T}}{1-v^2}
\int_{\rho_+}^{\rho_s} \frac{d\rho}{\rho'} \, 
\frac{ \big( 1+\vk_-^2\cosh^2\rho \big)(\cosh^2\rho -v^2(1-\vk_+^2 \sinh^2\rho) )  }{1+\vk_-^2\cosh^2\rho -\vk_+^2\sinh^2 \rho}
 \,,\\
S=& 
\hat{T} \int d\sigma \, g_{\psi\psi} \dot\psi 
% \nn\\=&
% \omega \hat{T} \int dy \, 
% \frac{\sinh^2\rho(1-\vk_+^2\sinh^2\rho )}{1+\vk_-^2\cosh^2\rho -\vk_+^2\sinh^2 \rho}
%  (1-v h_2')
%\nn\\=&
=
\frac{\omega \hat{T}}{1-v^2} 
\int_{\rho_+}^{\rho_s} \frac{d\rho}{\rho'} \,
\frac{\sinh^2\rho(1-\vk_+^2\sinh^2\rho )}{1+\vk_-^2\cosh^2\rho -\vk_+^2\sinh^2 \rho}
\bigg(1+\frac{v^2}{\omega}\vk_+\vk_-\bigg)
 \,,\\
J_2=& 
\hat{T} \int d\sigma \, g_{\phi\phi} \dot\phi
% \nn\\=&
% \Omega \hat{T} \int dy  
% \frac{\sin^2\theta(1+\vk_+^2 \sin^2\theta)}{1+\vk_-^2\cos^2\theta +\varkappa_+^2\sin^2\theta}
%\nn\\=&
=
\hat{T}
\int_{\rho_+}^{\rho_s} \frac{d\rho}{\rho'}
\,,
\end{align}
and $J_1=0$.
The lower end $\rho_+$ should be understood as $\rho_+=0$ if it takes a complex value.
The dispersion relation is expressed as
\begin{align}
   E-\frac{J_2}{\Omega} = \frac{S}{\omega} + K(\vk_+,\vk_-)  \,,
\end{align}
where the reminder function $K(\vk_+,\vk_-)$ is
\begin{align}
&  K(\vk_+,\vk_-) 
\nn\\=&
\frac{\hat{T}}{\omega(1-v^2)} \int dy \,
\frac{\sinh^2 \rho \big[\omega \vk_+^2 \cosh^2\rho +\omega(1+v^2 \vk_+^2) \vk_-^2 \cosh^2\rho
-v^2 \vk_+\vk_- (1-\vk_+^2 \sinh^2 \rho)   \big]}{1+\vk_-^2\cosh^2\rho -\vk_+^2\sinh^2 \rho}
\,,
\end{align}
which vanishes in the $\vk_\pm \rightarrow 0$ limit as it should, and
the dispersion relation of the undeformed case in \cite{Ryang:2006yq} is recovered.

\begin{figure}[tbh]
  \centering
  \includegraphics[scale=0.4]{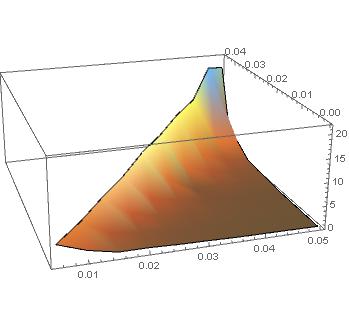}
\hspace{2em}
  \includegraphics[scale=0.7]{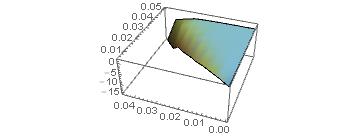} \\
  \includegraphics[scale=0.5]{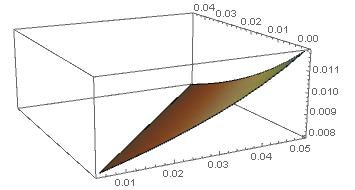}
\hspace{2em}
  \includegraphics[scale=0.5]{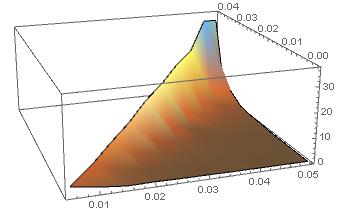}
 \put(-320,140){\makebox(0,0){$10^{-3} E/\hat{T}$}}
 \put(-140,140){\makebox(0,0){$10^{-3} S/\hat{T}$}}
 \put(-300,-10){\makebox(0,0){$10^{-3} J_2/\hat{T}$}}
 \put(-100,-10){\makebox(0,0){$10^{-3} K/\hat{T}$}}
% \put(-135,160){\makebox(0,0){\tiny $\vk_-=0$}}
% \put(-70,120){\makebox(0,0){\tiny $\vk_-=0.3$}}
% \put(-70,95){\makebox(0,0){\tiny $\vk_-=0$}}
%  \includegraphics[scale=0.7]{./spiky_omega_chg_small_omega.eps}
% \put(-193,140){\makebox(0,0){\tiny $\omega=0.9$}}
% \put(-130,160){\makebox(0,0){\tiny $\omega=0.7$}}
  \caption{The energy, the spins and the reminder function in the region (I),
for $0 \leq \vk_- \leq \vk_+ \leq 0.05$.
$\omega=1.0$ and $v=0.1$.
The scale of the vertical axes is divided by $1000$.}
  \label{fig:dispersion_relation}
\end{figure}

In the case of the region (I), the conserved charges are finite ($\rho_s$ works as a natural cutoff).
For example, for $\omega=1$, $v=0.1$, $\vk_-=0.1$ and $\vk_+=0.2$ (this is in the region (I),
\begin{align}
  E / \hat{T} = 322.691,
\qquad
  S / \hat{T} = -39.427,
\qquad
  J_2 / \hat{T}  = 63.829 \,,
\qquad
K / \hat{T} = 298.289 \,.
\end{align}
For small values of $\vk_\pm$, we plot the energy, the spins and the reminder function 
in Figure~\ref{fig:dispersion_relation}.
The reality condition \eqref{eq:reality_cond_omega} for $\omega=1.0$ and $v=0.1$
suggests that the real values exist about $\vk_\pm \leq 0.2$.

\section{PP-wave limit}
\label{sec:pp-wave-limit}

Following \cite{Ishizeki:2008tx,Losi:2010hr}, we will take the pp-wave limit.  First we perform the coordinate transformation in \eqref{eq:A3} and \eqref{eq:S3}
\begin{equation}
z=2\sqrt{2}e^{\rho_0-\rho},\qquad x_\pm = e^{\rho_0\mp \theta_0} (\psi \pm t)
\end{equation}
and then take following limits: send both $\rho_0,\theta_0 \to \infty$ but keep their difference finite such that $e^{\theta_0-\rho_0}\equiv 2\mu$.
We would also like to have a controllable way to incorporate the deformation.  A simple but nontrivial limit is to send $\vk_\pm \to 0$, but instead keep $\vk_\pm e^{\rho_0}$ finite.  
The metric after scaling becomes
\begin{equation}
ds^2_{pp} \simeq \frac{1}{z^2} [2dx_+dx_- -\mu^2 z^2 dx_+^2 -\frac{\mu^2}{z^2}\Delta^2 dx_+^2+dz^2],
\end{equation}
where $\Delta \equiv 2(\vk_+ - \vk_-)e^{2\rho_0}$.
We briefly look at two types of solutions in this background.

\paragraph{Moving straight string solution}

We consider the following ansatz \cite{Ishizeki:2008tx},
\begin{align}
x_+=\tau \,, \qquad
x_-= V \tau \,,
\qquad
z=z(\sigma) \,.
\end{align}
By looking at a point with constant $z$ (a point on the
world-sheet with a fixed $\sigma$), 
\begin{align}
  ds^2 
% =\frac{1}{z^2}
% \bigg(
% -\mu^2\bigg( z^2 + \frac{\Delta^2}{z^2} \bigg) d x_+^2
% +2 dx_+dx_- \bigg)
=\frac{1}{z^2}\bigg(
2V
-\mu^2\bigg( z^2 + \frac{\Delta^2}{z^2} \bigg)
 \bigg) d\tau^2 \,.
\end{align}
we find a condition for a part of the string not to travel faster than light;
namely, $z$ has to satisfy the condition,
\begin{align}
z^2 + \frac{\Delta^2}{z^2}  \geq \frac{2V}{\mu^2} \,,
\label{eq:cond_PP-straight}
\end{align}
and the string may not reach the boundary.

The Nambu-Goto action is
\begin{align}
  I
= - T \int d^2 \sigma \,
\frac{z'}{z}\sqrt{\mu^2 \bigg( z^2+\frac{\Delta^2}{z^2}\bigg) -2V }
\,,
\end{align}
where we use $z \geq 0$ and assume $z' \geq 0$; namely $z(\sigma)$ is
monotonic in $\sigma$.
With respect to the choice of the parameters, different ranges of $z$ are
allowed in the condition \eqref{eq:cond_PP-straight}:
\begin{itemize}
\item If $V^2 \leq \mu^4 \Delta^2$, arbitrary $z$ (and then $\sigma$)
  satisfies
the condition.
The solution is $z=\sigma$, $0\leq \sigma < \infty$,
and the string reaches the boundary.
It covers the above light-like solution.
Note that for $\Delta \neq 0$, the string can reach the boundary for
$V>0$.
\item If $V^2 > \mu^4 \Delta^2$, we have two branches:
one is $z=\sigma$, $\sigma_1 \leq \sigma < \infty$ with
\begin{align}
  \sigma_1 = \sqrt{\frac{V}{\mu^2}} \sqrt{1 + \sqrt{1-\frac{\mu^4
  \Delta^2}{V^2} }} \,,
\end{align}
which is similar to the straight string solution of
\cite{Ishizeki:2008tx}; a folded string with the spike not reaching
the boundary.
The other is $z=\sigma$, $0 \leq \sigma \leq \sigma_2$ with
\begin{align}
  \sigma_2 = \sqrt{\frac{V}{\mu^2}} \sqrt{1 - \sqrt{1-\frac{\mu^4
  \Delta^2}{V^2} }} \,,
\end{align}
which is a folded string solution that reaches to the boundary and it
turns back at a point in the bulk.
\end{itemize}
Now we focus on the solution with $\sigma_1 \leq \sigma \leq \infty$,
which is a generalization of the straight line solution.
Note that this case corresponds to a small $\Delta$.
The conserved charges are
\begin{align}
P_+ 
=&
%  \mu T \int_a^b \frac{d\sigma}{\sigma^2} \frac{\sigma^2 +
%        \frac{\Delta^2}{\sigma^2} - \frac{V}{\mu^2}}{\sqrt{\sigma^2 +
%        \frac{\Delta^2}{\sigma^2} - \frac{2V}{\mu^2}}}
% =
% \mu T \int_a^b \frac{d\sigma}{\sigma^3} \frac{\sigma^4 
% - \frac{V}{\mu^2} \sigma^2 + \Delta^2 }{\sqrt{\sigma^4 - \frac{2V}{\mu^2}\sigma^2+\Delta^2 }}
% \nn\\=&
\mu T \bigg[
-\frac{\sqrt{\sigma^4 - \frac{2V}{\mu^2}\sigma^2+\Delta^2 }}{2\sigma^2}
+\frac{1}{2}\log \bigg( -\frac{V}{\mu^2} + \sigma^2 +\sqrt{\sigma^4 - \frac{2V}{\mu^2}\sigma^2+\Delta^2 }  \bigg)
 \bigg]_a^b
\,,\\
P_-=&
%  -\frac{T}{\mu} \int_a^b \frac{d\sigma}{\sigma^2}
% \frac{1}{\sqrt{\sigma^2 +
%        \frac{\Delta^2}{\sigma^2} - \frac{2V}{\mu^2}}}
% =
% -\frac{T}{\mu} \int_a^b \frac{d\sigma}{\sigma}
% \frac{1}{\sqrt{\sigma^4 - \frac{2V}{\mu^2} \sigma^2 + \Delta^2}}
% \nn\\=&
\frac{T}{2\mu \Delta} \bigg[
\log \frac{2\Delta^2- \frac{2V}{\mu^2}\sigma^2 + 2\Delta\sqrt{\sigma^4
- \frac{2V}{\mu^2}\sigma^2 +       \Delta^2 } }{\sigma^2}
\bigg]_a^b
 \,,
\end{align}
where $a$ and $b$ represent the two ends of the string.
For $\sigma_1 \leq \sigma  <\infty$,
we introduce a large $R$ for a cutoff.
\begin{align}
  P_+ =& 
\mu T \bigg[ \frac{1}{2}\big( -1 + \log 2R^2 \big) + O(R^{-4})
-\frac{1}{2}\log \bigg( \sigma_1^2 - \frac{V}{\mu^2} \bigg) \bigg]
\label{eq:P+Spike1}
\,,\\
P_- =& 
% \frac{T}{2\mu\Delta} \bigg[
% \log \bigg( -\frac{2V}{\mu^2} + 2\Delta \bigg)
% - \log \bigg( \frac{2\Delta^2}{\sigma_1^2} - \frac{2V}{\mu^2} \bigg)
% \bigg]
% \nn\\=&
\frac{T}{2\mu\Delta} \bigg[
\log \bigg(1 -\frac{\Delta\mu^2}{V}  \bigg)
- \log \bigg(1- \frac{\mu^2\Delta^2}{V \sigma_1^2}  \bigg)
\bigg] \,.
\label{eq:P-Spike1}
\end{align}
A standard dictionary of the folded string solution in $AdS$ space reads\cite{Kruczenski:2008bs}
\begin{align}
  P_+=& P_t + P_\theta =  E-S \,, \qquad
P_-= -P_t + P_\theta = -(E+S) \,.
\end{align}
Then
\begin{align}
  S=& - \frac{1}{2}(P_+ + P_-)
\nn\\=&
\frac{|P_-|}{2}
+ \frac{\mu T }{4} \bigg[ 1
+ \log \bigg( \frac{\Delta}{2R^2} \coth \bigg[ \frac{2\mu \Delta |P_-| }{T}\bigg]
\sqrt{1- \tanh^2 \frac{2\mu \Delta |P_-| }{T}}  \bigg) \bigg]
\,.
\end{align}
This relation can be inverted in the large $S$ limit.
We first consider a small $\Delta$ limit and take inversion,
\begin{align}
|P_-| =& 2S +\frac{\mu T}{2} \bigg(-1+ \ln \frac{4\mu R^2 |P_-|}{T}  \bigg)
+ \frac{\mu^2 \Delta^2 |P_-|^2}{3T} +\mathcal{O}(\Delta^4)
\nn\\=&
2S +\frac{\mu T}{2} \bigg( \ln S + \ln \frac{8\mu R^2}{T} -1 \bigg)
+\frac{\mu^2 T^2}{8S} \bigg(\ln S + \ln \frac{8\mu R^2}{T} -1 \bigg)
+\mathcal{O}(S^{-2})
\nn\\&
+ \Delta^2
\bigg[
\frac{4\mu^3 S^2}{3T}
+\frac{2 \mu^4 S}{3} \bigg(
\ln S +  \ln \frac{8 \mu R^2}{T} -\frac{1}{2} 
\bigg) +\mathcal{O}(\ln S) \bigg]
\nn\\&
+\mathcal{O}(\Delta^4) \,.
\end{align}
Thus, the dispersion relation is
 expressed as
\begin{align}
E-S=&  P_+= \frac{\mu T}{2} \bigg(-1+ \ln |P_-| + \ln \frac{4\mu R^2}{T} \bigg)
+ \frac{\mu^3 |P_-|^2}{3T}\Delta^2 
+\mathcal{O}(\Delta^4)
\nn\\=&
\frac{\mu T}{2} \bigg( \ln S + \ln \frac{8\mu R^2}{T} -1 \bigg)
+\frac{\mu^2 T^2}{8S} \bigg(\ln S + \ln \frac{8\mu R^2}{T} -1 \bigg)
\nn\\&
+ \Delta^2
\bigg[
\frac{4\mu^3 S^2}{3T}
+\frac{2 \mu^4 S}{3} \bigg(
\ln S +  \ln \frac{8 \mu R^2}{T} -\frac{1}{2} 
\bigg) \bigg]
\nn\\&
+\cdots \,.
\end{align}
In the large-$S$ limit, $\Delta$ correction terms become dominant
and the standard relation $E-S \propto \ln S$ is spoiled.
It may suggest that two limits ($S\rightarrow \infty$ and $\Delta \rightarrow 0$)
are not interchangeable.

\paragraph{Periodic spike solutions}

We next consider the ansatz for the 
periodic spike solution \cite{Ishizeki:2008tx},
\begin{align}
  x_+=\tau \,,
\qquad x_-=\sigma \,,
\qquad
z=z(\xi) \,.
\end{align}
Here
\begin{align}
  \xi= \tau- \frac{1}{\eta_0^2} \sigma = -\frac{v-1}{\sqrt{2}} \big(
  x- vt \big) \,, \qquad
  \eta_0^2= \frac{v-1}{v+1}
\end{align}
and $v>1$.
A point with fixed $z$ is therefore traveling in $x$ direction with
speed $v$.
By using $\partial_\tau = \partial_\xi$ and $\partial_\sigma =
-\frac{1}{\eta_0^2}\partial_\xi$,
the equation of motion for $x_-$ becomes
\begin{align}
  0=& \partial_\sigma \bigg[ \frac{x_-'+ \dot{z} z'}{z^2 F} \bigg]
 -\partial_\tau \bigg[ \frac{z^{\prime 2}}{z^2 F} \bigg]
= -\partial_\xi \bigg[ \frac{1}{\eta_0^2 z^2 F} \bigg]
\end{align}
with
\begin{align}
  F= \sqrt{1- \frac{(\partial_\xi z)^2}{\eta_0^2} \bigg\{ 2- \frac{\mu^2}{\eta_0^2}
  \bigg( z^2 + \frac{\Delta^2}{z^2} \bigg) \bigg\} } \,.
\end{align}
This can be integrated as
\begin{align}
  \partial_\xi z =&
\frac{\eta_0^2}{\mu z^2}
\sqrt{\frac{z_0^4-z^4}{z^2+\frac{\Delta^2}{z^2}-z_1^2}}
=
\frac{\eta_0^2}{\mu z}
\sqrt{\frac{z_0^4-z^4}{z^4-z_1^2 z^2 + \Delta^2}} 
\end{align}
with $z_1=\sqrt{2} \eta_0/\mu$ and $z_0$ is a constant of integration, $z^2 F = z_0^2$.

Here, we look at $z$ equation of motion,
\begin{align}
  0=& \partial_\xi \bigg[
\frac{\partial_\xi z}{\eta_0^2 z^2 F} \bigg( 2- \frac{\mu^2}{\eta_0^2} \bigg(z^2 + \frac{\Delta^2}{z^2} \bigg) \bigg)
\bigg]
+\frac{2F}{z^3} 
- \frac{\mu^2 (\partial_\xi z)^2 }{\eta_0^4 z^2 F} \bigg( z - \frac{\Delta^2}{z^3} \bigg)
\,.
\end{align}
Now we apply the following relations that are from $x_-$ equation of
motion (and its $\xi$ derivative),
\begin{align}
\label{eq:x-eom2}
  z^2 F =& z_0^2 \,, \qquad
\bigg( 2- \frac{\mu^2}{\eta_0^2} \bigg(z^2 + \frac{\Delta^2}{z^2} \bigg) \bigg)
\partial_\xi^2 z 
= \frac{2\eta_0^2 z_0^4}{z^5} + \frac{\mu^2 (\partial_\xi z)^2}{\eta_0^2} \bigg( z - \frac{\Delta^2}{z^3} \bigg) \,,
\end{align}
and then
\begin{align}
  0=& 
\frac{\partial_\xi^2 z}{\eta_0^2 z_0^2} \bigg( 2- \frac{\mu^2}{\eta_0^2} \bigg(z^2 + \frac{\Delta^2}{z^2} \bigg) \bigg)
-\frac{2 \mu^2 (\partial_\xi z)^2}{\eta_0^4 z_0^2} \bigg( z - \frac{\Delta^2}{z^3} \bigg)
%\nn\\&
-\frac{2 z_0^2}{z^5}
+ \frac{ \mu^2 (\partial_\xi z)^2}{\eta_0^4 z_0^2} \bigg( z - \frac{\Delta^2}{z^3} \bigg)
 \,,
\end{align}
where in the first equality the first relation of \eqref{eq:x-eom2} is used
and in the last the second used.
Namely, $z$ equation of motion is also satisfied.

The turning point is given by $z=z_0$.
The position of the spike is
\begin{align}
  z^2 = \frac{1}{2} \big( z_1^2 \pm \sqrt{ z_1^4 - 4 \Delta^2 } \big)
= \frac{\eta_0^2}{\mu^2} \bigg( 1 \pm \sqrt{1- \frac{\mu^4
  \Delta^2}{\eta_0^4}} \bigg) \,,
\end{align}
for $\eta_0^2 > \mu^2 \Delta$.
If $\eta_0^2 \leq \mu^2 \Delta$, the denominator inside the square
root does not take zero, and the position of the spike is $z=0$;
namely the spike reaches to the boundary.
In this case, we do not investigate the dispersion relation in detail, 
but we present the expressions of 
the conserved charges,
\begin{align}
  P_+ =& -2 \frac{\eta_0^2 T}{z_0^2} \int_{\tilde{z}_0}^{\tilde{z}_1} \frac{dz}{\partial_\xi z}
\bigg[1- \frac{(\partial_\xi z)^2}{\eta_0^2} \bigg\{ 1- \frac{2}{z_1^2}
  \bigg( z^2 + \frac{\Delta^2}{z^2} \bigg) \bigg\}  \bigg]
\nn\\=&
-\frac{2\mu T}{z_0^2} \int_{\tilde{z}_0}^{\tilde{z}_1} dz \bigg[
z \sqrt{\frac{z^4-z_1^2 z^2 + \Delta^2}{z_0^4-z^4}}
- \frac{z_1^2}{2z} 
\sqrt{\frac{z_0^4-z^4}{z^4-z_1^2 z^2 + \Delta^2}} 
\bigg\{ 1- \frac{2}{z_1^2}
  \bigg( z^2 + \frac{\Delta^2}{z^2} \bigg) \bigg\}
\bigg]
\,,\\
P_-=& 2 \frac{\eta_0^2 T}{z_0^2} \int_{\tilde{z}_0}^{\tilde{z}_1} dz \, \frac{1}{\eta_0^4} \partial_\xi z
%\nn\\=&
=\frac{2 T}{\mu z_0^2} \int_{\tilde{z}_0}^{\tilde{z}_1} dz \, 
\frac{1}{z} \sqrt{\frac{z_0^4-z^4}{z^4-z_1^2 z^2 + \Delta^2}} \,,
\end{align}
where ${\tilde{z}_0}$ and ${\tilde{z}_1}$ are the turning points of the string solution
and the factor two comes from the configuration being folded.
It is easy to check that they come back to (33) and (34) of \cite{Ishizeki:2008tx}
in the $\Delta \rightarrow 0$ limit.

% To study spiky string solution, we assume the embedding ansatz
% \begin{equation}
% u=\tau, \quad v=\sigma, \quad z=z(\sigma)
% \end{equation}
% The induced Polyakov action reads
% \begin{equation}
% S =-\frac{\sqrt{\lambda}}{4\pi}\int d\tau d\sigma \frac{1}{z^2}[2\partial_a u \partial^a v - (\mu^2z^2+\frac{\mu^2\Delta^2}{z^2})\partial_a u \partial^a u +\partial_a z \partial^a z] 
% \end{equation}
% or  Nambu-Goto action (after ansatz)
% \begin{equation}
% S =-\frac{\sqrt{\lambda}}{2\pi}\int d\tau d\sigma \frac{1}{z^2}\sqrt{(z^2\mu^2+\frac{\mu^2\Delta^2}{z^2})z^{\prime 2}+1}
% \end{equation}
% The equation of motion can be derived from this action
% \begin{equation}
% z^\prime = \sqrt{(\mu^2 z^2 + \frac{\mu^2\Delta^2}{z^2})^{-1}(\frac{z_0^4}{z^4}-1)}
% \end{equation}

% The conserved momentum flows are
% \begin{eqnarray}
% P_u &&= \frac{\sqrt{\lambda}}{2\pi} \int d\sigma (\mu^2 \dot{u}+\frac{\mu^2\Delta^2}{z^4}\dot{u}-\frac{\dot{v}}{z^2}),\nonumber\\
% P_v &&= \frac{\sqrt{\lambda}}{2\pi} \int d\sigma \frac{\dot{u}}{z^2}
% \end{eqnarray}

\section{Conclusion}
\label{sec:conlusion}

In this paper, we have considered the classical string solutions in the 
two parameter deformation of AdS space constructed by Hoare \cite{Hoare:2014oua}.
We first observe that in the fast spinning limit, the string Hamiltonian coincides with
a spin chain Hamiltonian in a small deformation parameter limit.
Therefore, the integrability may be preserved in this background at least in the case of small
deformation.
We further construct the giant magnon solutions and the hanging string solutions.
They are obtained as a generalization of the one-parameter deformation case.
We also derive the expression of the conserved charges.
It however turns out that the dispersion relation takes a fairly complicated form
in the two parameter deformation; we then consider a perturbative expansion of the 
dispersion relation.
In this geometry, there appears the surface of singularity whose location 
is determined by the two deformation parameters.
In the hanging string case, there appear several types of solutions; hung from the singular surface, stretching between the boundary and the singular surface, or reaching to the center of AdS space.
We also numerically evaluated the energy and the spins in this case.
Finally, we consider the PP-wave limit of this background.
The PP-wave background is characterized by a parameter which is essentially the difference
of two deformation parameter.
When this parameter is small, we can evaluate the dispersion relation for the moving straight string solution. We also briefly looked at the existence of periodic spike solutions.
In the appendix, we briefly discuss the PP-wave limit near the singular surface and find a similar structure that 
appears in the PP-wave limit near the boundary.
This would imply another clue for the similarity between the theories near 
the singular surface and the AdS boundary.

The solutions obtained in this paper are generalization of the solutions in the one parameter deformed background.
Although the results including two deformation parameter is complicated and not so illuminating,
we can consider some future directions.
First, we may calculate various physical quantities with these classical solutions; the holographic entanglement
entropy or the complexity from the Wheeler-de Wit patch are some examples.
At least in the limit of the parameters discusses in this paper, we will be able to evaluate these values
and observe how their behavior changes due to the deformations.
The dual gauge theory corresponding to this geometry is still unclear.
We also hope that some hints on the dual gauge theory side are obtained through
further study of the classical solutions presented in this paper.

\section*{Acknowledgments}
We are grateful to Dr. Shogo Kuwakino for his contribution in the early stage of this project.  This work is supported in parts by the Taiwan's Ministry of Science and Technology (grant No. 106-2112-M-033-007-MY3 for WYW and grant No.~107-2112-M-033-008 and 108-2112-M-033-003 for SK) and the National Center for Theoretical Science.

%%%%%%%%%%%%%%%%%%%%% Appendix %%%%%%%%%%%%%%%%%%%%%%%%%%
\appendix
%%%%%%%%%%

\section{Two-parameter deformation of $AdS_3 \times S^3$ geometry}
\label{sec:two-param-deform}

In this appendix, we present a briefly summary of the two-parameter $q$-deformed $AdS_3 \times S^3$
geometry presented by Hoare\cite{Hoare:2014oua} and the singular surface of this geometry.
AdS part and the sphere part metrics are given by
\begin{align}
  ds^2=& ds_{A_3}^2 + ds_{S^3}^2 \,,\\
\label{eq:A3_0}
ds_{A_3}^2 =&
\frac{1}{1+\vk_-^2(1+\tilde{\rho})^2-\vk_+^2\tilde{\rho}^2}
\nn\\& \times
\bigg[
\frac{d\tilde{\rho}^2}{1+\tilde{\rho}^2}
-(1+\tilde{\rho}^2)\big( 1+\vk_-^2(1+\tilde{\rho}^2) \big) dt^2
+\tilde{\rho}^2(1-\vk_+^2\tilde{\rho}^2) d\psi^2
+2\varkappa_+\varkappa_-\tilde{\rho}^2(1+\tilde{\rho}^2) dtd\psi
\bigg]\,,\\
\label{eq:S3_0}
ds_{S^3}^2=&
\frac{1}{1+\vk_-^2(1-r^2)+\varkappa_+^2r^2}
\nn\\& \times
\bigg[
\frac{dr^2}{1-r^2}
+(1-r^2)\big( 1+\vk_-^2(1-r^2) \big) d\varphi^2
+r^2(1+\vk_+^2 r^2)d\phi^2
+2\vk_-\vk_+r^2(1-r^2) d\varphi d\phi
\bigg] \,.
\end{align}
$\vk_\pm$ are two deformation parameters of the metric.
There is $U(1)^4$ isometry that corresponds to constant shifts in
$t, \psi, \varphi$, and $\phi$, and the corresponding conserved charges
are $E, S, J_1$ and $J_2$.
Note that this metric has a $\mathbf{Z}_2$ symmetry \cite{Hoare:2014oua},
$r \rightarrow \sqrt{1-r^2}$,
$\varphi \leftrightarrow \phi$, and
$\vk_+ \leftrightarrow \vk_-$
(equivalent to
$\theta \rightarrow \frac{\pi}{2}-\theta$
$\phi_1 \rightarrow \phi_1$,
$\phi_2 \rightarrow -\phi_2$, and
$\vk_+ \leftrightarrow \vk_-$),
and the we can assume that
$\vk \geq \vk_-$ in this paper.

By changing the coordinates,
$\tilde{\rho} \rightarrow \sinh \rho$ and
$r \rightarrow \sin \theta$,
we obtain
\begin{align}
\label{eq:A3}
ds_{A_3}^2 
=&
\frac{1}{1+\vk_-^2\cosh^2\rho -\vk_+^2\sinh^2 \rho}
\bigg[
d\rho^2
-\cosh^2\rho \big( 1+\vk_-^2\cosh^2\rho \big) dt^2
\nn\\& \hskip6em
+\sinh^2\rho(1-\vk_+^2\sinh^2\rho ) d\psi^2
+2\varkappa_+\varkappa_-\sinh^2\rho \cosh^2\rho dtd\psi
\bigg]\,,\\
\label{eq:S3}
 ds_{S^3}^2
=&
\frac{1}{1+\vk_-^2\cos^2\theta +\varkappa_+^2\sin^2\theta}
\bigg[
d\theta^2
+\cos^2\theta\big( 1+\vk_-^2\cos^2\theta \big) d\varphi^2
\nn\\& \hskip6em
+\sin^2\theta(1+\vk_+^2 \sin^2\theta)d\phi^2
+2\vk_-\vk_+ \sin^2\theta \cos^2\theta d\varphi d\phi
\bigg] \,.
\end{align}
% In the analysis, we often refer to the metric components,
% $ds^2 = g_{\mu\nu} dx^\mu dx^\nu$ where
% \begin{align}
% g_{\rho\rho}=& \frac{1}{1+\vk_-^2\cosh^2\rho -\vk_+^2\sinh^2 \rho} \,,\qquad
% g_{\theta\theta}= \frac{1}{1+\vk_-^2\cos^2\theta +\varkappa_+^2\sin^2\theta}
% \,,\\
%   g_{tt}=& \big[ -\cosh^2\rho \big( 1+\vk_-^2\cosh^2\rho \big) \big] g_{\rho\rho}
% \,,\quad
% g_{\psi\psi}= \big[ \sinh^2\rho(1-\vk_+^2\sinh^2\rho ) \big] g_{\rho\rho}
% \,,\\
% g_{t\psi}=& \big[\varkappa_+\varkappa_-\sinh^2\rho \cosh^2\rho \big] g_{\rho\rho}
% \,,\\
% g_{\phi\phi}=& \big[ \sin^2\theta(1+\vk_+^2 \sin^2\theta) \big] g_{\theta\theta}
% \,,\quad
% g_{\varphi\varphi}= \big[ \cos^2\theta\big( 1+\vk_-^2\cos^2\theta \big)\big] g_{\theta\theta}
% \,,\\
% g_{\varphi\phi}=& \big[\vk_-\vk_+ \sin^2\theta \cos^2\theta \big] g_{\theta\theta}
% \,.
% \end{align}

The AdS part of the metric has a singularity (or rather a surface of singularity) at
$\rho=\rho_s$ with $\rho_s$ being a solution of
\begin{align}
  \cosh^2 \rho_s = \frac{1+\vk_+^2}{\vk_+^2-\vk_-^2} \,.
\end{align}
It can be checked that this is the curvature singularity as in the case of one-parameter
deformed geometry.
From a point in the bulk, the singular surface can be reached in a finite coordinate time $t$ but it takes
infinite affine time as also in the one-parameter deformation case,
The coordinate time from the center to the singular surface reads
\begin{align}
  t =& \int_0^{\rho_s} \frac{d\rho}{\cosh\rho \sqrt{ 1+\vk_-^2\cosh^2\rho }}
%\nn\\=&
=
\text{arctan} \, 
\bigg[
\frac{ \sinh \rho_s}{\sqrt{1+ \vk_-^2 \cosh^2 \rho_s}}
 \bigg]
= \text{arctan} \, \frac{1}{\vk_+}
\,.
\end{align}
Interestingly, this result is independent of the second deformation parameter
$\vk_-$ and precisely agrees with that of one-parameter deformation case
\cite{Kameyama:2014vma}.
Especially, in a limit, $\vk_- \rightarrow \vk_+$ ($\vk_+ \geq \vk_-$), 
the singular surface is pushed to infinity, $\rho_s \rightarrow \infty$,
but the coordinate time is still $t=\text{arccot}\, \vk_+$.
Under the undeformed limit $\vk_+ \rightarrow 0$, we have $t=\pi/2$.

Next we consider the affine parameter $t_A$.
In the massless condition,
$0=p^2= G^{tt} p_t p_t + G^{\rho\rho} p_\rho p_\rho$ with
$E = -p_t$ and $p_\rho = G_{\rho\rho} \frac{d\rho}{dt_A}$ leads to
\begin{align}
-  \frac{1+\vk_-^2\cosh^2\rho -\vk_+^2\sinh^2 \rho}{\cosh^2\rho \big( 1+\vk_-^2\cosh^2\rho \big)}
E^2
+ \frac{1}{1+\vk_-^2\cosh^2\rho -\vk_+^2\sinh^2 \rho} \bigg( \frac{d\rho }{dt_A} \bigg)^2
=0 \,.
\end{align}
Thus,
\begin{align}
  t_A =&
 \int_0^{\rho_s} \frac{d\rho}{E} \frac{\cosh\rho \sqrt{ 1+\vk_-^2\cosh^2\rho }}{1+\vk_-^2\cosh^2\rho -\vk_+^2\sinh^2 \rho}
% \nn\\=&
% -\frac{1}{2E(\vk_+^2-\vk_-^2)}\bigg[
% 2\vk_- \ln \bigg( \frac{\sqrt{1+\vk_-^2}}{\sqrt{1+\vk_-^2 \cosh^2\rho}-\vk_-\sinh \rho} \bigg)
% +\vk_+ \ln \bigg(\frac{\sqrt{1+\vk_-^2 \cosh^2\rho}-\vk_+\sinh \rho}{\sqrt{1+\vk_-^2 \cosh^2\rho}+\vk_+\sinh \rho} \bigg)
% \bigg]_{\rho \rightarrow \rho_s}
 \nn\\=&
 -\frac{1}{2E(\vk_+^2-\vk_-^2)}\bigg[
 2\vk_- \ln \bigg( \frac{\sqrt{1+\vk_-^2 \cosh^2\rho}+\vk_-\sinh \rho }{\sqrt{1+\vk_-^2}} \bigg)
 \nn\\& \hskip6em
+\vk_+ \ln \bigg(\frac{1+\vk_-^2 \cosh^2\rho-\vk_+^2 \sinh^2 \rho}{\big(1+\vk_-^2 \cosh^2\rho-\vk_+^2 \sinh^2 \rho \big)+2\vk_+\sinh \rho \sqrt{1+\vk_-^2 \cosh^2\rho} } \bigg)
\bigg]_{\rho \rightarrow \rho_s}
\nn\\ &
\rightarrow \infty \,.
\end{align}
Thus, it takes infinite affine time to reach the singular surface.
Note that in the $\vk_-$ limit, we recover the one parameter deformation case
\begin{align}
  t_A=& \frac{1}{E \vk_+} \text{arctanh}\, \big(\vk_+ \sinh \rho \big) \bigg|_{\rho \rightarrow \rho_s}
=\infty \,.
\end{align}

\paragraph{New coordinates}

By following \cite{Kameyama:2014vma}, we may define a new coordinate
$\chi$ as
\begin{align}
  \cosh \chi =& \sqrt{-g_{tt}} = \frac{\cosh\rho \sqrt{1+\vk_-^2 \cosh^2\rho}}{\sqrt{1+\vk_-^2 \cosh^2\rho - \vk_+^2 \sinh^2\rho}} \,,
\end{align}
which covers $0 \leq \chi \leq \infty$ for $0 \leq \rho \leq \rho_s$.
With this new variable, the AdS part of the metric can be expressed as
\begin{align}
g_{tt} dt^2 =&
-\cosh^2\chi dt^2 \,,\\
  g_{\psi\psi} d\psi^2 =&
-\frac{1}{2\vk_+^2 \vk_-^2}
\bigg[
(\vk_+^2+\vk_-^2+2 \vk_+^2\vk_-^2) + (\vk_+^4+\vk_-^4) \cosh^2\chi
\nn\\& \hskip6em 
- (\vk_+^2 + \vk_-^2) \sqrt{f_1(\vk_+,\vk_-;\chi)} 
\bigg] d\psi^2
% \nn\\ =&
% \bigg( \frac{\sinh^2\chi}{1+\vk_+^2 \cosh^2\chi} + O(\varkappa_-^2) \bigg) d\psi^2 
\,,\\
g_{t\psi} dtd\psi =&
\frac{1+(\vk_+^2+\vk_-^2)\cosh^2\chi - \sqrt{f_1(\vk_+,\vk_-;\chi)}}{2 \vk_+\vk_-} dtd\psi
% \nn\\=&
% \bigg(\frac{\vk_+^2\vk_-^2 \cosh^2\chi \sinh^2\chi}{1+\vk_+^2 \cosh^2\chi}+ O(\varkappa_-^4) 
% \bigg) dtd\psi 
\,,\\
g_{\rho\rho} d\rho^2 =&
\frac{1+(\vk_+^2+\vk_-^2)\cosh^2\chi + \sqrt{f_1(\vk_+,\vk_-;\chi)}}{2 f_1(\vk_+,\vk_-;\chi)}
d\chi^2
% \nn\\=&
% \bigg( \frac{1}{1+\vk_+^2 \cosh^2\chi} + O(\vk_-^2) \bigg) d\chi^2 
\,,\\
f_1(\vk_+,\vk_-;\chi) =&
4\vk_-^2(1+\vk_+^2)\cosh^2\chi + \big( 1+(\vk_+^2-\vk_-^2)\cosh^2\chi \big)^2 \,.
\end{align}
With these new coordinates, we may revisit the PP-wave limit.
The standard pp-wave limit corresponds to zooming up the near boundary region
of the AdS space, which is a part seen by a fast rotating string.
We now consider a similar limit in this new coordinate, which corresponds to
zooming up near the singular surface.

The definition of the new coordinates and the limit are the same,
where we use $\chi$ instead of $\rho$ that covers inside region bounded by the
singular surface $\rho=\rho_s$:
\begin{align}
  z=2\sqrt{2} e^{\chi_0-\chi} \,,
\qquad
x_\pm = e^{\chi_0 \mp \theta_0}(\psi \pm t) \,.
\end{align}
Taking the following limit:
\begin{align}
  \rho_0, \theta_0 \rightarrow \infty \,,
\qquad
e^{\theta_0-\chi_0}=2\mu=\text{fixed.}
\end{align}
Plugging these in to the aforementioned new metric,
and taking the limit $\vk_\pm \rightarrow 0$ with $\vk_\pm e^{2\rho_0}=\mu_\pm$ kept finite
($\Delta = 2(\mu_+ - \mu_-)$),
we find that the reduced metric is
\begin{align}
    ds^2
\simeq &
\frac{1}{z^2}
\bigg[
dz^2
-\mu^2 z^2 d x_+^2
+2 dx_+dx_- 
-\frac{\mu^2 \Delta^2}{z^2} dx_+^2 
\bigg]
 \,.
\end{align}
In this simplified limit, this is equivalent to the PP-wave limit in the original coordinates
as discussed in Section \ref{sec:pp-wave-limit}.
It might suggest that the properties near the AdS boundary and the singular surface are
quite similar at least in this limit.

%%%%%%%%%%%%%%%%%%%%%%%%%%%%%%%%%%%%%%%%%%%%%%%%%%%%%%%%%%%%%%%


\begin{thebibliography}{99}
\bibitem{Maldacena:1997re}
  J.~M.~Maldacena,
  ``The Large N limit of superconformal field theories and supergravity,''
  Int.\ J.\ Theor.\ Phys.\  {\bf 38} (1999) 1113
   [Adv.\ Theor.\ Math.\ Phys.\  {\bf 2} (1998) 231]
  doi:10.1023/A:1026654312961
  [hep-th/9711200].

\bibitem{integ_big_review}
For a thourough review, 
N. Beisert et al.,
``Review of AdS/CFT Integrability: An Overview,''
Lett. Math. Phys. \textbf{99}, 3 (2012), [arXiv:1012.3982]

\bibitem{Swanson:2007dh}
  I.~Swanson,
  ``A Review of integrable deformations in AdS/CFT,''
  Mod.\ Phys.\ Lett.\ A {\bf 22} (2007) 915
  doi:10.1142/S0217732307023614
  [arXiv:0705.2844 [hep-th]].


%\cite{Hoare:2014oua}
\bibitem{Hoare:2014oua} 
  B.~Hoare,
  ``Towards a two-parameter q-deformation of AdS$_3 \times S^3 \times M^4$ superstrings,''
  Nucl.\ Phys.\ B {\bf 891}, 259 (2015)
  [arXiv:1411.1266 [hep-th]].
  %%CITATION = ARXIV:1411.1266;%%
  %13 citations counted in INSPIRE as of 30 Aug 2015

\bibitem{Arutyunov:2013ega}
  G.~Arutyunov, R.~Borsato and S.~Frolov,
  ``S-matrix for strings on $\eta$-deformed AdS5 x S5,''
  JHEP {\bf 1404} (2014) 002
  doi:10.1007/JHEP04(2014)002
  [arXiv:1312.3542 [hep-th]].


\bibitem{Cherednik:1981df}
  I.~V.~Cherednik,
  ``Relativistically Invariant Quasiclassical Limits of Integrable Two-dimensional Quantum Models,''
  Theor.\ Math.\ Phys.\  {\bf 47} (1981) 422
   [Teor.\ Mat.\ Fiz.\  {\bf 47} (1981) 225].
  doi:10.1007/BF01086395

%\cite{Kruczenski:2003gt}
\bibitem{Kruczenski:2003gt} 
  M.~Kruczenski,
  ``Spin chains and string theory,''
  Phys.\ Rev.\ Lett.\  {\bf 93}, 161602 (2004)
  [hep-th/0311203].
  %%CITATION = HEP-TH/0311203;%%
  %250 citations counted in INSPIRE as of 31 Aug 2015

%\cite{Kameyama:2014bua}
\bibitem{Kameyama:2014bua} 
  T.~Kameyama and K.~Yoshida,
  ``Anisotropic Landau-Lifshitz sigma models from $q$-deformed AdS$_{5} \times$ S$^{5}$ superstrings,''
  JHEP {\bf 1408}, 110 (2014)
  [arXiv:1405.4467 [hep-th]].
  %%CITATION = ARXIV:1405.4467;%%
  %10 citations counted in INSPIRE as of 06 Dec 2014

%\cite{Wen:2006fw}
% \bibitem{Wen:2006fw} 
%   W.~Y.~Wen,
%   ``Spin chain from marginally deformed AdS(3) x S**3,''
%   Phys.\ Rev.\ D {\bf 75}, 067901 (2007)
%   [hep-th/0610147].
%   %%CITATION = HEP-TH/0610147;%%
%   %8 citations counted in INSPIRE as of 31 Aug 2015

\bibitem{Wen:2006fw}
  W.~Y.~Wen,
  ``Spin chain from marginally deformed $AdS_3 \times S^3$,''
  Phys.\ Rev.\ D {\bf 75} (2007) 067901
  [hep-th/0610147].


%\cite{Ishizeki:2008tx}
\bibitem{Ishizeki:2008tx} 
  R.~Ishizeki, M.~Kruczenski, A.~Tirziu and A.~A.~Tseytlin,
  ``Spiky strings in AdS(3) x S1 and their AdS-pp-wave limits,''
  Phys.\ Rev.\ D {\bf 79}, 026006 (2009)
  [arXiv:0812.2431 [hep-th]].
  %%CITATION = ARXIV:0812.2431;%%
  %17 citations counted in INSPIRE as of 31 Aug 2015

%\cite{Losi:2010hr}
\bibitem{Losi:2010hr} 
  M.~Losi,
  ``Spiky strings and the AdS/CFT correspondence,''
  arXiv:1109.5401 [hep-th].
  %%CITATION = ARXIV:1109.5401;%%
  %3 citations counted in INSPIRE as of 31 Aug 2015

%\cite{Arutyunov:2014cda}
\bibitem{Arutyunov:2014cda} 
  G.~Arutyunov and D.~Medina-Rincon,
  ``Deformed Neumann model from spinning strings on ($AdS_5 \times S^5$)$_\eta$,''
  JHEP {\bf 1410}, 50 (2014)
  [arXiv:1406.2536 [hep-th]].
  %%CITATION = ARXIV:1406.2536;%%
  %9 citations counted in INSPIRE as of 06 Dec 2014

%\cite{Khouchen:2014kaa}
\bibitem{Khouchen:2014kaa} 
  M.~Khouchen and J.~Kluson,
  ``Giant Magnon on Deformed AdS(3)xS(3),''
  Phys.\ Rev.\ D {\bf 90}, 066001 (2014)
  [arXiv:1405.5017 [hep-th]].
  %%CITATION = ARXIV:1405.5017;%%
  %7 citations counted in INSPIRE as of 06 Dec 2014

%\cite{Ahn:2014aqa}
\bibitem{Ahn:2014aqa} 
  C.~Ahn and P.~Bozhilov,
  ``Finite-size giant magnons on $\eta$-deformed $AdS_5 \times S^5$,''
  Phys.\ Lett.\ B {\bf 737}, 293 (2014)
  [arXiv:1406.0628 [hep-th]].
  %%CITATION = ARXIV:1406.0628;%%
  %8 citations counted in INSPIRE as of 06 Dec 2014

%\cite{Banerjee:2014bca}
\bibitem{Banerjee:2014bca} 
  A.~Banerjee and K.~L.~Panigrahi,
  ``On the rotating and oscillating strings in (AdS$_{3}$  x S$^{3}$)$_{\kappa}$,''
  JHEP {\bf 1409}, 048 (2014)
  [arXiv:1406.3642 [hep-th]].
  %%CITATION = ARXIV:1406.3642;%%
  %6 citations counted in INSPIRE as of 06 Dec 2014

\bibitem{Dai:2014twa}
  S.~H.~Dai, S.~Kuwakino and W.~Y.~Wen,
  ``Spin chains and classical strings in rotating Rindler-AdS space,''
  JHEP {\bf 1404} (2014) 018
  [arXiv:1401.6915 [hep-th]].
\bibitem{Ryang:2006yq}
  S.~Ryang,
  ``Three-spin giant magnons in AdS(5) x S**5,''
  JHEP {\bf 0612} (2006) 043
  [hep-th/0610037].

\bibitem{Kruczenski:2008bs}
  M.~Kruczenski and A.~A.~Tseytlin,
  ``Spiky strings, light-like Wilson loops and pp-wave anomaly,''
  Phys.\ Rev.\ D {\bf 77} (2008) 126005
  doi:10.1103/PhysRevD.77.126005
  [arXiv:0802.2039 [hep-th]].

\bibitem{Kameyama:2014vma}
  T.~Kameyama and K.~Yoshida,
  ``A new coordinate system for $q$-deformed AdS$_{5} \times$ S$^5$ and classical string solutions,''
  J.\ Phys.\ A {\bf 48} (2015) 7,  075401
  [arXiv:1408.2189 [hep-th]].

\end{thebibliography}
\end{document}